\documentclass[a4paper,10pt]{article}

\pdfoutput=1
\usepackage{amsfonts}
\usepackage{amsmath}
\usepackage{amssymb}
\usepackage[english]{babel}
\usepackage[makeroom]{cancel}
\usepackage[font={small},labelfont=bf]{caption}
\usepackage{empheq}
\usepackage{floatrow}
\usepackage[margin=0.80in]{geometry}
\usepackage{graphicx}
\usepackage{hyperref}
\usepackage[utf8]{inputenc}
\usepackage{lscape}
\usepackage{makecell}
\usepackage{mathtools}
\usepackage{multirow}
\usepackage{parskip}
\usepackage{pdfpages}
\usepackage{placeins}
\usepackage[outercaption]{sidecap}
\usepackage{siunitx}
\usepackage{tikzsymbols}
\usepackage{titling}
\usepackage[normalem]{ulem}
\usepackage{xcolor}
\usepackage[square, numbers, comma, sort&compress]{natbib}
\usepackage[affil-it]{authblk}

\setcellgapes{4pt}

\newcommand{\stkout}[1]{\ifmmode\text{\sout{\ensuremath{#1}}}\else\sout{#1}\fi}

\newcommand{\fin}{F_{\text{in}}}
\newcommand{\xM}{x_{\text{M}}}
\newcommand{\finc}{F_{\text{in}}^{\text{c}}}
\newcommand{\ch}{\text{c}}

\DeclareSIUnit\Molar{\textsc{M}}
\DeclareSIUnit\bacteria{\text{bacteria}}
\DeclareSIUnit\OD{\textsc{OD}}

\makeatletter
\g@addto@macro\bfseries{\boldmath}
\makeatother

\usepackage{tocloft}
\title{Hydrodynamic flow and concentration gradients in the gut enhance neutral bacterial diversity }
\date{\today}
\author{Darka Labavi{\'c}$^1$, Claude Loverdo$^{1,*,\dagger}$ and Anne-Florence Bitbol$^{2,3,*,\dagger}$}
\affil{$^1$Sorbonne Universit{\'e}, CNRS, Institut de Biologie Paris-Seine, Laboratoire Jean Perrin (UMR 8237), F-75005 Paris, France, \\$^2$Institute of Bioengineering, School of Life Sciences, École Polytechnique Fédérale de Lausanne (EPFL), CH-1015 Lausanne, Switzerland, \\$^3$SIB Swiss Institute of Bioinformatics, CH-1015 Lausanne, Switzerland,\\$^*$Equal contribution, \\$^\dagger$E-mail addresses: claude.loverdo@sorbonne-universite.fr; anne-florence.bitbol@epfl.ch.}

\begin{document}

\maketitle

\begin{abstract}
The gut microbiota features important genetic diversity, and the specific spatial features of the gut may shape evolution within this environment.
We investigate the fixation probability of neutral bacterial mutants within a minimal model of the gut that includes hydrodynamic flow and resulting gradients of food and bacterial concentrations. We find that this fixation probability is substantially increased compared to an equivalent well-mixed system, in the regime where the profiles of food and bacterial concentration are strongly spatially-dependent. Fixation probability then becomes independent of total population size. We show that our results can be rationalized by introducing an active population, which consists of those bacteria that are actively consuming food and dividing. The active population size yields an effective population size for neutral mutant fixation probability in the gut. 
\end{abstract}

\section*{Significance statement}
The human body harbors numerous and diverse bacteria, the vast majority of which are located in the gut. These bacteria can mutate and evolve within the gut, which is their natural environment. This can have important public health implications, e.g. when gut bacteria evolve antibiotic resistance. The gut features specific characteristics, including hydrodynamic flow and resulting  gradients of food and bacterial concentrations. How do these characteristics impact the evolution and diversity of gut bacteria? We demonstrate that they can substantially increase the probability that neutral mutants reach high proportions and eventually take over the population. This is because only a fraction of gut bacteria is actively dividing. Thus, the specific environment of the gut enhances neutral bacterial diversity. 

\section*{Introduction}

In the human body, bacteria are approximately as numerous as human cells, and about 99\% of these bacteria are located in the digestive tract \cite{sender2016}. The gut microbiota is very diverse, and collectively harbors more genes than there are human genes \cite{Turnbaugh2007}. One source of this genetic diversity is evolution occurring within the gut, which is the natural environment of these bacteria. Such evolution can have important public health implications, as the gut can constitute a reservoir of antibiotic resistance, both in humans and in farm animals \cite{Landers2012}. How does the environment in the gut affect the evolution of bacteria? A crucial feature of the gut is the flow of its contents along its main axis, and the associated gradients of concentration of food and bacteria. Going downstream along this axis, food is first ingested, then simple nutrients are absorbed by the body, next more complex molecules are broken down by bacteria, and eventually what remains of the food exits the system, together with many bacteria, which make up from a quarter to half of fecal mass \cite{rose2015characterization}. These features yield a very particular spatial structure that can impact the evolution of bacteria.

Evolutionary models that investigate population spatial structure generally consider discrete patches of population with migrations between them, and the same environment in each of them~\cite{Wright31,Kimura64,Maruyama70,Maruyama74,Nagylaki80,Slatkin81,Barton93,Whitlock97,Whitlock03}. Complex spatial structures are investigated through models on graphs where each individual~\cite{Lieberman05,Kaveh15, Hindersin15, Pattni15} or each patch of population~\cite{Houchmandzadeh11,Houchmandzadeh13,Constable14,MarrecPreprint} occupies a node of the graph. Population structure can impact the rapidity of adaption~\cite{Gordo06,Perfeito08,Kryazhimskiy12,Korona94,Bitbol14,Nahum15,France19} because local competition can allow the maintenance of larger genetic diversity. In simple population structures where migration is symmetric between patches~\cite{Wright31,Kimura64}, the fixation probability of a mutant is unaffected by population structure~\cite{Maruyama70,Maruyama74}, unless extinctions of patches occur~\cite{Barton93}. However, more complex population structures with asymmetric migrations can impact the fixation probabilities of beneficial and deleterious mutants~\cite{Whitlock03,Lieberman05,MarrecPreprint}. In the case of the gut, the flow can be viewed as yielding asymmetric migrations, but the system is continuous. In large-scale turbulent systems, hydrodynamic flow has been shown to strongly impact fixation probabilities and fixation times~\citep{Perlekar10,Pigolotti12,Plummer19}. In addition, environmental gradients, e.g. of antibiotic concentration, can strongly impact evolution \cite{Zhang11,Greulich2012, Hermsen12, Baym2016}. How do population structure, hydrodynamic flow and gradients shape the evolution of bacteria in the gut microbiota?

Here, we propose a minimal model of evolution of bacteria in the gut. Because most bacteria in the human digestive tract are located in the bulk of the colon lumen \cite{donaldson2016,sender2016}, we focus on this compartment. Since most bacteria in the digestive tract have no self-motility \cite{kurokawa2007,turnbaugh2009}, we consider that they are carried passively with the digesta. The motion of the digesta is complex, but it was shown in Refs.~\cite{Cremer2016,Cremer2017} that it can be approximated as a one-dimensional flow with net velocity and effective diffusion representing mixing. Within this model of the gut that includes hydrodynamic flow and resulting gradients of food and bacterial concentrations, we ask how the fixation probability of a neutral mutant compares with that in an equivalent well-mixed chemostat. We find that the structure of the gut can increase this fixation probability, specifically in the regime where the profiles of food and bacterial concentration are strongly spatially-dependent. In this regime, fixation probability becomes independent of total population size, in stark contrast with a well-mixed population, where fixation probability is inversely proportional to total population size~\cite{Moran58,Ewens79}. We show that this behavior can be understood by introducing the notion of active population, which corresponds to the fraction of the bacterial population that is actively consuming food and dividing. 

\section*{Model and methods}

Because the majority of bacteria in the human digestive tract are in the colon \cite{donaldson2016}, we focus on this compartment. Within the colon, there are marked differences between bacteria associated to mucus and bacteria in the digesta, i.e. in the bulk of the colon lumen \cite{donaldson2016}. The latter constitute the majority of bacteria in the colon. Indeed, the surface area of the large intestine, including its folds, is about 2 square meters \cite{helander2014}, while the mucus layer is about $100-300~\mu$m thick \cite{cone2009}, and typically comprises a few $10^{8}$ bacteria per milliliter in healthy samples \cite{Swidsinski2005}, which leads to an order of magnitude of $10^{11}$ bacteria associated to mucus. This number is small compared to the total colon content, which is around $10^{14}$ bacteria \cite{sender2016}. 
Since mucus-associated bacteria constitute a small minority in the colon, and since their spatial structure and migration patterns are not well characterized, we focus on the bacteria present in the bulk of the colon lumen, and do not model the mucus layer. Henceforth, we refer to the colon lumen by ``gut" for simplicity. 

The dynamics of wild-type and mutant bacteria and food in the gut is described through three concentration fields, of food $F$, wild-type bacteria $B$ and mutant bacteria $M$, based on the description of the coupled dynamics of food and bacteria (without mutants) developed in Ref.~\cite{Cremer2016}. The gut is represented by a tube of length $L$ and cross-section with surface area $S$ (Figure~\ref{fig1}A). In addition to this cylindrical symmetry, we neglect radial variations, and are left with a one dimensional model along the $x$ axis, specifically a segment of length $L$. We assume a constant inflow of nutrients at the entrance of this gut segment and no inflow of bacteria. At the exit of the gut, we assume that there is a free outflow of both nutrients and bacteria. The dynamics is affected by the constant flow velocity $v$, by the mixing due to different mechanisms e.g. peristaltic movement, which is modelled by effective diffusion with diffusion coefficient $D$, and by the harvesting of the food by bacteria, which is described by a Hill-type function with Monod constant $k$, and is coupled to their growth which has maximal rate $r$. This leads to the following coupled partial differential equations:
\begin{subequations}\label{eq:PDEs}
  \begin{align}
	\frac{\partial F}{\partial t} & =  D \frac{\partial^2F}{\partial x^2} - v\frac{\partial F}{\partial x} - \frac{r}{\alpha}\frac{(B+M) F}{k+F},\label{eq:model_F} \\
	\frac{\partial B}{\partial t} & =  D \frac{\partial^2B}{\partial x^2} - v\frac{\partial B}{\partial x} + r \frac{B F}{k+F},\label{eq:model_B} \\
	\frac{\partial M}{\partial t} & =  D \frac{\partial^2M}{\partial x^2} - v\frac{\partial M}{\partial x} + r \frac{MF}{k+F},\label{eq:model_M} 
\end{align}
\end{subequations}
with boundary conditions
\begin{subequations}\label{eq:PDEs_BC}
	 \begin{align}
	-D \frac{\partial [F;B;M]}{\partial x}(x=0) & + v[F;B;M](x=0)  =  [v\fin       ;0;0]\,,\\
	-D \frac{\partial [F;B;M]}{\partial x}(x=L) & =  [0;0;0],
\end{align}
\end{subequations}
where $[F;B;M]$ denotes a vector.
Here $v\fin $ is the food inflow at the entrance of the gut segment, while $\alpha$ denotes the yield of the conversion from food to bacteria. Note that there is zero inflow of bacteria, in agreement with observations that
bacterial concentration in the smaller intestine is orders of magnitude smaller~\cite{donaldson2016,Cremer2016,Cremer2017}. The boundary conditions at $x=L$ cancel the diffusive flux, corresponding to free outflow toward the downstream part of the colon.

In our study of the fate of mutants appearing in the gut, initial conditions are
\begin{subequations}\label{eq:PDEs_IC}
	 \begin{align}
	F(t=0,x) & =  F^*(x),\\
	B(t=0,x) & =  B^*(x),\\
	M(t=0,x) & =  \begin{cases}
M_0, |x-\xM|\leq\Delta x/2,\\
0, |x-\xM|>\Delta x/2,
\end{cases}
\end{align}
\end{subequations}
where $F^*$ and $B^*$ represent the steady state of system~\ref{eq:PDEs} without mutant bacteria, while $\xM\in(0,L)$ is the position in the gut where the mutant appears, while $\Delta x$ is a short length, taken equal to the spatial discrete step in our numerical resolutions,  and $M_0\ll B(\xM)$ is the initial local concentration of mutant at this location. In practice, $M_0$ is set through $N_\text{M}=M_0 S \Delta x$, where $S$ is the surface area of the section of the gut, so that the total number $N_\text{M}$ of mutants introduced in the system is always the same, and our results do not depend on $\Delta x$ as long as it is small compared to the length scale over which concentrations vary.

The partial differential equations in Eqs.~\ref{eq:PDEs} with boundary conditions in Eqs.~\ref{eq:PDEs_BC} and initial conditions in Eqs.~\ref{eq:PDEs_IC} were solved numerically (Supplementary material Section~\ref{SM:num_meth}, and code publicly available~\cite{codeZenodo}).

\section*{Results}

\subsection*{Spatial dependence of the steady-state bacterial concentration}

Our aim is to study the fate of neutral mutants appearing in the gut, starting from initial conditions where the concentrations of food and wild-type bacteria are at steady state (see Eqs.~\ref{eq:PDEs_IC}). Therefore, we start by describing the steady-state profiles of food and wild-type bacteria in the mutant-free gut.

\begin{figure}[h!]
\includegraphics[width=\textwidth]{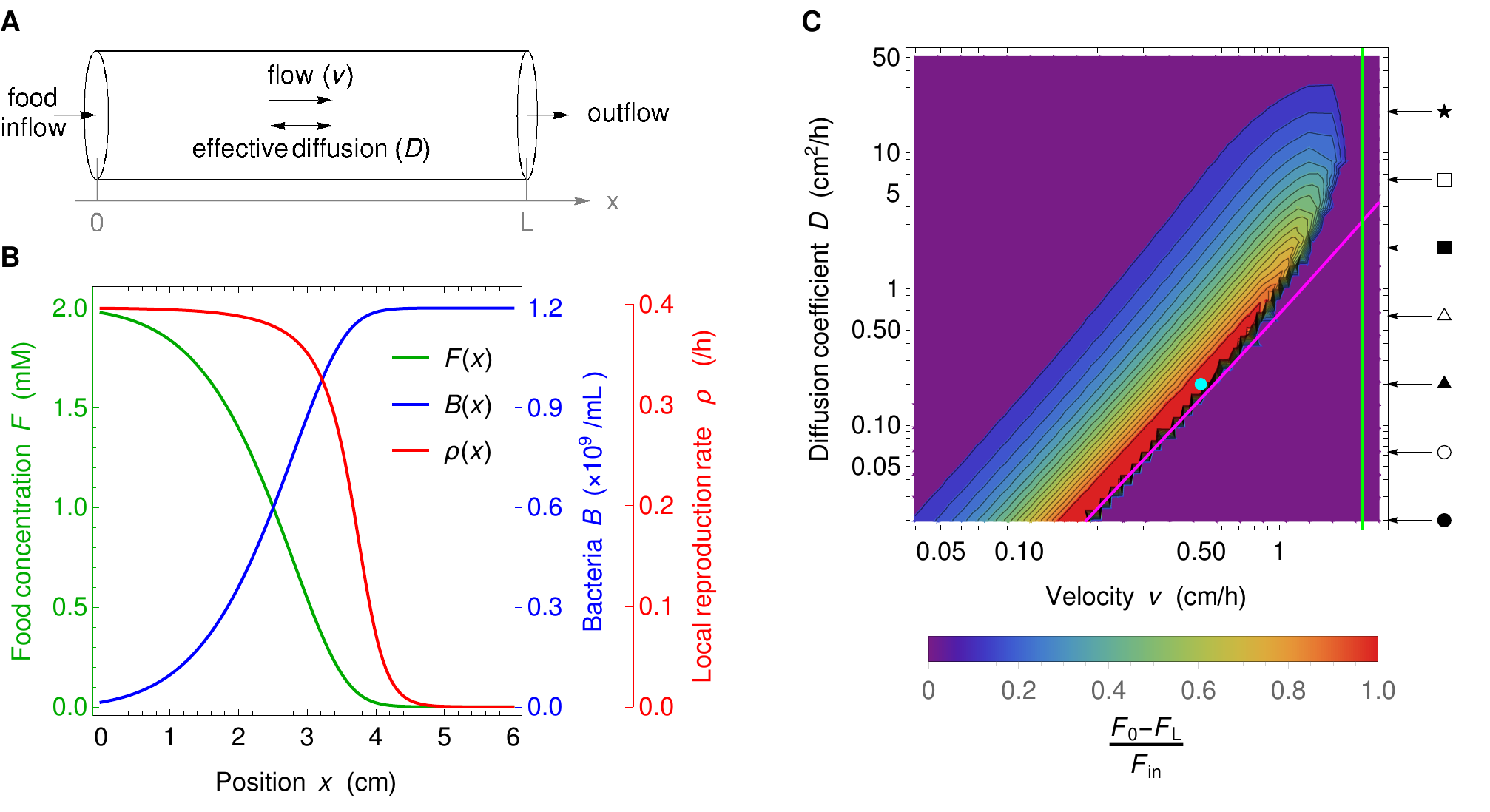}
\caption{\textbf{Model of the gut and associated spatial gradients.} \textbf{A}: Schematic representation of the gut model investigated. We consider a cylinder with length $L$ and neglect concentration variations in the radial direction, thus simplifying the system to one dimension along the $x$ axis. Transport is modeled as flow with constant velocity $v$ and effective diffusion coefficient $D$.  At the upstream boundary $x=0$ we consider constant food inflow, and no bacteria inflow, while at the downstream boundary $x=L$ we consider zero diffusive outflow.
\textbf{B}: Concentration $F$ of food, amount $B$ of bacteria, and reproduction rate $\rho$ of bacteria versus the coordinate $x$ along the gut. Curves are numerical solutions of Eq.~\ref{eq:PDEs} for $D=$ \SI[per-mode = symbol]{0.2}{\centi\meter\squared\per\hour}, $v=$ \SI[per-mode = symbol]{0.5}{\centi\meter\per\hour}, $k=$ \SI[per-mode = symbol]{0.1}{\milli\Molar}, $r=$ \SI{0.42}{\per\hour}, $v\:\fin=$ \SI[per-mode = symbol]{1}{\milli\Molar\centi\meter\per\hour}, $\alpha=$ \SI[per-mode = symbol]{6.13e8}{\bacteria \per\milli\Molar} and no mutant bacteria. The section area, $S$, is taken to be \SI{1}{\centi\meter\squared} in the entire paper, and the length $L$ is 6 cm, as in the mini-gut of Ref.~\citep{Cremer2016}, in the main text, but is varied in the Supplementary material (see Section~\ref{SM:param}). The parameters are chosen such that they fall in range of parameters compared to the experiments in~\citep{Cremer2016} and that the concentration profile is dependent of the spatial coordinate. The depicted concentrations represent the state of the system after numerically integrating partial differential Eqs.~\ref{eq:PDEs} for time $t=$\SI{500}{\hour} which is sufficient to reach the steady state.
\textbf{C}: Heat map of the level of spatial dependence of the concentration profiles, quantified by $[F(0)-F(L)]/\fin$, versus $v$ and $D$. High values of $[F(0)-F(L)]/\fin$ (red) mean strong gradients in the gut. Magenta and green lines represent washout limits, resp. $D=v^2(k/\fin       +1)/(4r)$ and $v=rL/(k/\fin+1)$. Below the magenta line and on the right side of the green line, there are no bacteria in the gut at steady state, while in the purple region on the top left hand side, the system is well-mixed, leading to an almost uniform but non-zero concentration of bacteria in the gut. Parameter values (except $v$ and $D$) are the same as in panel B. The values of $v$ and $D$ used in panel B are indicated by a circular cyan marker. 
Arrows and symbols on the right hand-side of the heat map indicate the diffusion coefficient values employed in Figure~\ref{fig4} with the same symbols.}\label{fig1}
\end{figure}

Steady-state solutions of the spatial model described by Eqs.~\ref{eq:PDEs} can strongly depend on the spatial coordinate $x$ for some values of flow velocity $v$, and effective diffusion constant $D$, as exemplified by Figure~\ref{fig1}B. Such strong spatial dependence is relevant in the ascending colon~\citep{Cremer2017}, which is our focus here. These strongly spatial profiles resemble Fisher waves, and indeed, the steady-state equation describing our system has the same form as the equation satisfied by a traveling wave in a Fisher-KPP equation~\citep{Fisher37,KPP37,MurrayBook}. However, the velocity $v$ is here an imposed parameter, in contrast to a traveling wave velocity. Moreover, the nontrivial stationary solutions satisfying the boundary conditions Eqs.~\ref{eq:PDEs_BC} are in a different parameter regime compared to Fisher waves (see Supplementary material Section~\ref{secFisher}). We quantify the spatial dependence of the concentration profiles through the difference between food concentration at the entrance and at the exit of the gut, normalized by the incoming food concentration $\fin$, namely $[F(0)-F(L)]/\fin$. A heat map of this quantity is depicted in the $(v,D)-$parameter space in Figure~\ref{fig1}C. We observe diverse levels of spatial dependence, ranging from strongly spatial profiles to quasi-flat ones, where the system is almost well-mixed and resembles a chemostat, or where bacteria are washed out by the flow~\citep{Cremer2016,Cremer2017} (see Figure~\ref{fig:Food_profiles} for examples of concentration profiles across these regimes). There are two washout limits here. First, for large  diffusion coefficients, if the flow timescale is smaller than the replication timescale, bacteria exit the system before reproducing. Second, for small diffusion coefficients, on the timescale of one replication, if the characteristic length of flow is larger than that of diffusion, bacteria are washed out (see Supplementary material Section~\ref{SM:washout} and Figure~\ref{fig:WOL}).

To compare our spatial system to a well-mixed one, we consider a chemostat~\cite{Herbert1956} with the same total number of bacterial reproductions $ N_{\text{R}}$ per unit time as in the spatial system, which is
\begin{equation}\label{eq:Reprrate}
  N_{\text{R}} =S \int_0^L  B(x)\rho(x) dx ,
\end{equation}
where $S$ is the surface area of the gut section, while $\rho(x)$ is the reproduction rate of bacteria, which can be expressed using food concentration as in Eqs.~\ref{eq:PDEs}:
\begin{equation}\label{eq:Reprrate2}
\rho(x)=r\frac{F(x)}{k+F(x)}.
\end{equation}
This reproduction rate strongly depends on the spatial coordinate in the spatial regime of the concentration profiles, see Figure~\ref{fig1}B. Once the total number of reproductions is matched, it is possible to impose an additional matching condition, and we consider three possibilities for it in the Supplementary material (see Section~\ref{SM:chemostat}). These matching conditions allow to set the parameters characterizing the chemostat matching the spatial system, namely its dilution rate, food inflow and volume. In all cases, we observe that matching chemostats feature extreme values for some of these parameters (see Figure~\ref{fig:chem_param}), which arises from the very small outflow of food in the spatial system (see Supplementary material Section~\ref{SM:chemostat}). These results emphasize that the large intestine is a highly efficient system for converting unabsorbed nutrients into bacteria.

\newpage

\subsection*{Dynamics and fate of neutral mutants appearing in the gut}

Let us now consider neutral mutants that spontaneously appear in the gut at steady state. Mutants may appear at any position along the gut, which can feature strong spatial heterogeneities (see Figure~\ref{fig1}). How does the initial position of these mutants affect their dynamics and their steady-state concentration?

The initial local concentration of mutants is assumed to be much smaller than that of the wild type at the position $\xM$ where the mutants appear (see Eqs.~\ref{eq:PDEs_IC}), as we aim to describe the fate of a single mutant or a few mutants, but in the framework of the continuous description of the gut.
The early dynamics of mutant concentration is governed by the fluid dynamics in the gut. Indeed, the position $x$ with the highest mutant concentration at a given time $t$ initially follows the $x=\xM+vt$ line, while the time $t$ for which the mutant concentration is maximal for a given position $x$ initially follows the $t=(x-\xM)^2/(2D)$ curves (see Supplementary material, Figure~\ref{fig:MutDynam}). This is consistent with the infinite space solution of the diffusion equation obtained from Eqs.~\ref{eq:PDEs} when ignoring reproduction. Hence, transport by convective and diffusive flow allows the early spread of the mutants in the gut. Afterwards, coupling with the reproduction term and the boundary conditions yields more complex dynamics.

Because neutral mutant concentration satisfies the same partial differential equation as wild-type bacteria concentration (see Eqs.~\ref{eq:PDEs}), the steady-state concentrations of mutant and wild type bacteria satisfy $M(x)/B(x)=C$, where $C$ depends on the initial conditions but not on $x$. In other words, the steady-state concentration profile of neutral mutants versus position $x$ along the gut is the same as for wild-type bacteria, but with an overall rescaling. The magnitude of this rescaling (i.e. the value of $C$) depends on the initial mutant quantity, and on the position $\xM$ where mutant bacteria appear. The latter dependence on $\xM$ is strong in the regime where spatial dependence is strong in the mutant-free system (see Figure~\ref{fig1}C), as shown in Figure~\ref{fig2}A and Figures~\ref{fig:M(x,xM)} and~\ref{fig:MB(x,xM)} in the Supplementary material. If the number of mutants that appear is held constant, then mutants make up a much larger steady-state fraction of bacteria if they appeared close to the entrance of the gut than if they appeared close to its exit, because they have more opportunity to spread and divide in the gut.

\begin{figure}[ht!]
\includegraphics[width=\textwidth]{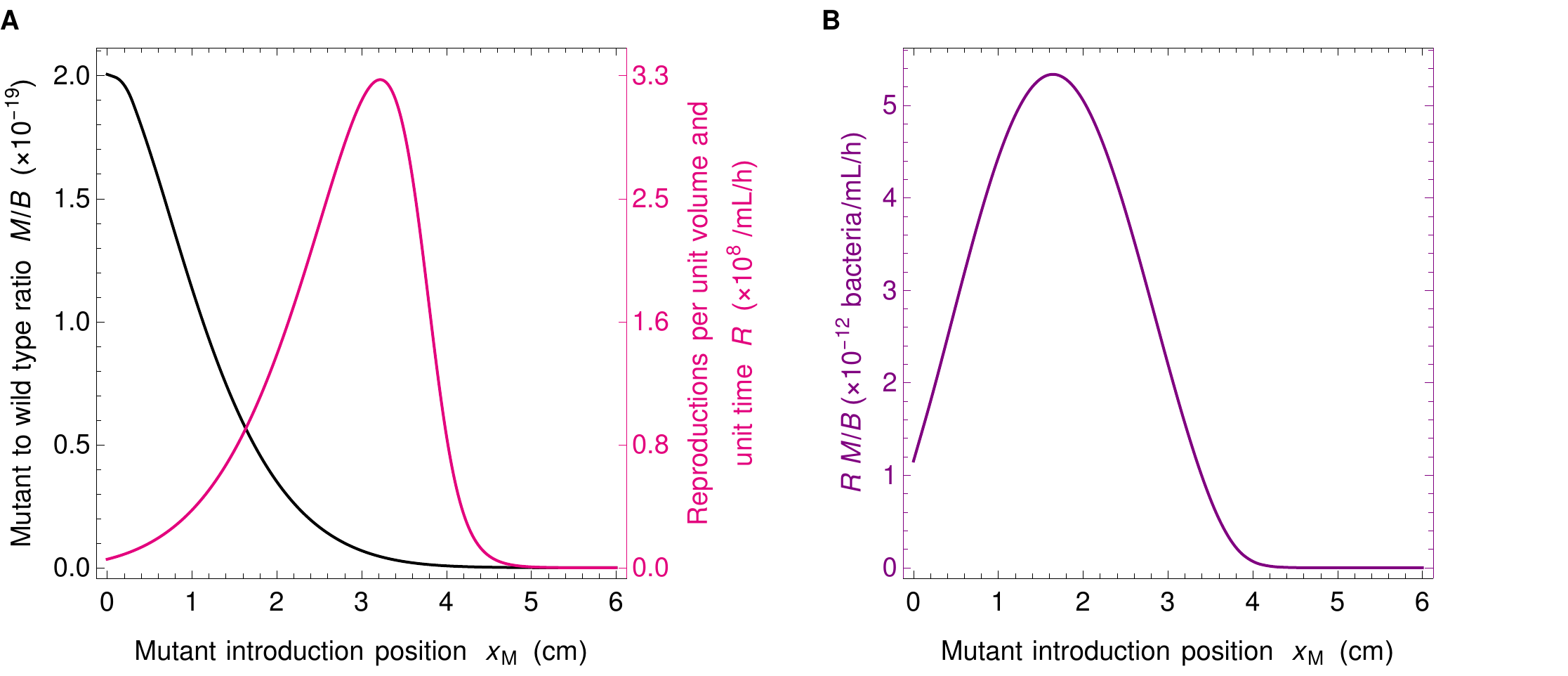}
\caption{\textbf{Fate of neutral mutants appearing at various locations in the gut.} \textbf{A}: Steady-state ratio $M/B$ of mutant to wild-type bacteria concentrations, and number of reproduction events $R$ per unit volume and unit time versus position $\xM$ of the mutant introduction. The ratio $M/B$ yields the fixation probability of a mutant that appears at a given position $\xM$ in the system. As mutants generally appear upon division, the appearance of new mutants is proportional to $R$, which thus also matters for the overall likelihood that a mutant appears and fixes. Mutants are introduced at numerical integration time $t=$ \SI{500}{\hour}, when steady state is reached. 
\textbf{B}: Product of the ratio $M/B$ and the number $R$ of reproductions per unit volume and unit time versus $\xM$. This quantity yields the fixation probability of a mutant that appears proportionally to reproduction rate.
Parameter values are the same as in Figure~\ref{fig1}B, and $F$ and $B$ are initially at steady state as in Figure~\ref{fig1}B, while mutants are introduced locally by using the initial condition in  Eqs.~\ref{eq:PDEs_IC}, with a total number $N_M=$ \SI{3.33e-11} of mutants introduced in the system. }\label{fig2}
\end{figure}

In our deterministic continuous description, bacterial species or strains coexist forever (except in the washout case where they are all wiped away), reflected by the fact that $M/B$ is nonzero at steady state. However, the fate of individual mutants is in fact affected by demographic fluctuations known as genetic drift~\cite{Ewens79}, so that in a finite system, mutants eventually either take over the population or disappear. Here, on a short time scale, mutants either reach deterministic steady state in coexistence with the wild-type, or they get extinct stochastically. If they reach steady state, then, on a longer time scale, proportional to population size~\cite{Ewens79}, one of the two types takes over. What is the probability that a mutant lineage that has reached steady state then fixes in the population? In a well-mixed system, the fixation probability of a neutral mutant is given by the ratio of the number of mutants to the total number of individuals~\cite{Ewens79}. In our gut model, the steady-state ratio $M/B$ is independent of $x$ in the deterministic limit (note that throughout we have $M\ll B$ so that here $M/(M+B)\approx M/B$, and we only discuss $M/B$). Moreover, in Eqs.~\ref{eq:PDEs}, the only non-linearity in the evolution of $B$ and $M$ comes from the dependence of $F$ on $B$ and $M$. Here, since we introduce a very small amount of mutants, $M_0\ll B$, when $B$ is at stationary state, and since the overall bacterial population is very large, $F$ remains almost constant through the evolution of $M$, which entails that the equations for $B$ and $M$ are then approximately linear. Because in the linear case, the equations on averages across replicates of a stochastic system coincide with those of the deterministic large-size limit \cite{gustafsson2013can}, the fixation probability of neutral mutants in the stochastic case is given by the deterministic steady-state ratio $M/(M+B)\approx M/B$. 
In section~\ref{sec:stochsim} of the Supplementary material, we provide a validation of our deterministic analysis by stochastic simulations. Fig.~\ref{fig:fig2_stoch} demonstrates the good agreement between the two descriptions regarding the fate of neutral mutants appearing at various locations in the gut. In particular, it confirms that the deterministic steady-state ratio $M/B$ yields the mutant fixation probability, for each given mutant introduction position $\xM$. 
Given the dependence of the ratio $M/B$ on the initial position $\xM$ of the mutants (see Figure~\ref{fig2}A and discussion above), mutants appearing close to the entrance of the gut are much more likely to fix than those appearing close to its exit, in the regime with strong spatial dependence (see Figure~\ref{fig1}C).

Where in the gut do the mutants that fix originate? To address this question, we need to account for the apparition of mutants as well as for their fixation. Assume that mutations occur upon division, which is the case for replication errors. Then, mutants appear at a position $\xM$ proportionally to the local number 
\begin{equation}
    R(\xM)=B(\xM)\rho(\xM) \label{eq:Reprrate3}
\end{equation}
of reproduction events per unit volume and unit time (where the reproduction rate $\rho$ is given by Eq.~\ref{eq:Reprrate2}). This number is small close to the exit of the gut because food is exhausted, but it is also small close to its entrance because bacteria are scarce, and it features a maximum at an intermediate location (see Figure~\ref{fig2}A). What ultimately sets the location where mutants that fix tend to originate is the product of $M/B$ and $R$, whose dependence on the mutant initial position $\xM$ is depicted in Figure~\ref{fig2}B. It features a strong spatial dependence, with a maximum at an intermediate position in the gut. In the Supplementary material section~\ref{SM:maxR}, we study $R$ and $R\,M/B$ for various parameter values, and show that this maximum of $R\,M/B$ at an intermediate position in the gut is obtained robustly in the regime with strong spatial dependence (see Figure~\ref{fig:maxR}). Furthermore, Fig.~\ref{fig:fig2_stoch} demonstrates the good agreement of the results obtained in our deterministic model with those from stochastic simulations.

\subsection*{Spatial structure in the gut increase the fixation probability of neutral mutants}

What is the overall probability $\mathcal{F}$ that neutral mutants fix in the gut,
averaged over their possible positions of origin? It can be expressed as the  integral over all possible initial mutant locations $\xM$ of the fixation probability given $\xM$, multiplied by the probability that the mutant originates at this location $\xM$:
\begin{equation}\label{eq:Fixprob}
    \mathcal{F} = \frac{\int_0^L R(\xM)\frac{M(\xM)}{B(\xM)}d\xM}{\int_0^L R(\xM)d\xM}.
\end{equation}

\begin{figure}[h!]
\includegraphics[width=.6\textwidth]{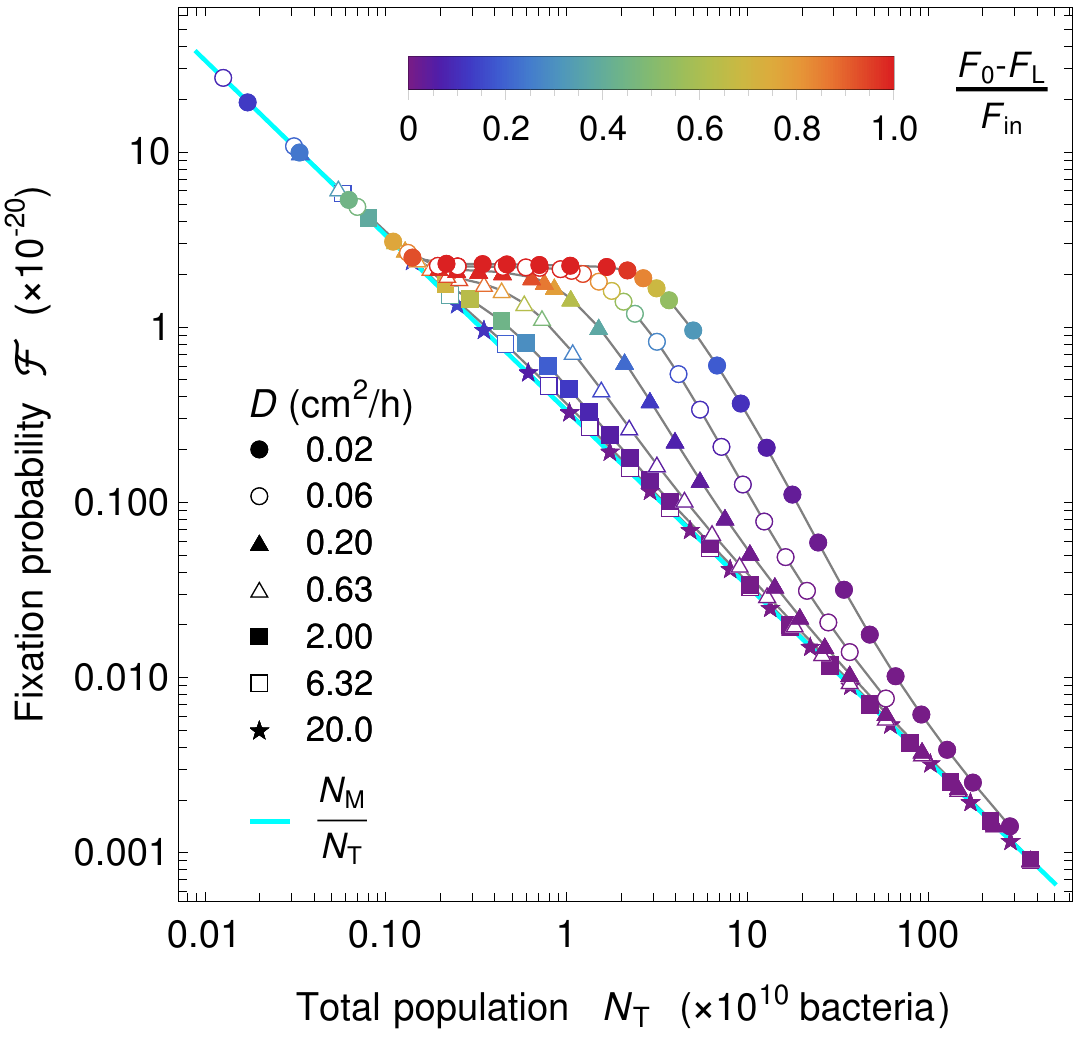}
\caption{
\textbf{Fate of a neutral mutant versus population size in the gut.} Fixation probability $\mathcal{F}$ of mutants appearing proportionally to reproduction rate is shown versus total population size $N_{\text{T}}$, for different diffusion coefficients $D$ (see corresponding markers on the right hand-side of the heat map in Fig.~\ref{fig1}C). Markers are colored by the level of spatial dependence of the concentration profiles, quantified by $[F(0)-F(L)]/\fin$ as in Fig.~\ref{fig1}C. For strong spatial dependence (red), a plateau is observed, evidencing a large difference with the well-mixed expectation $\mathcal{F}=N_{\text{M}}/N_{\text{T}}$.
For each value of $D$, $v$ is varied while keeping $v\:\fin$ constant, and fixation probability is calculated from Eq.~\ref{eq:Fixprob} and total population by integrating the sum of the mutant and the wild type in the total space (volume).  
Parameter values: $D\in [0.02,20.0]$ \SI[per-mode = symbol]{}{\centi\meter\squared\per\hour}, $v\in [0.001, 2.4]$ \SI[per-mode = symbol]{}{\centi\meter\per\hour}, $k=$ \SI[per-mode = symbol]{0.1}{\milli\Molar}, $r=$ \SI{0.42}{\per\hour}, $v\:\fin =$ \SI[per-mode = symbol]{1}{\milli\Molar\centi\meter\per\hour}, and $\alpha=$ \SI[per-mode = symbol]{6.13e8}{\bacteria \per\milli\Molar} and initial conditions as in Figure~\ref{fig2}.}\label{fig3}
\end{figure}

How is the overall fixation probability $\mathcal{F}$ of a neutral mutant affected by the spatial dependence of food and bacterial concentrations in the gut? To address this question, Figure~\ref{fig3} depicts $\mathcal{F}$ versus total population size $N_{\text{T}}=S\int_0^L B(x)dx$ for different velocities $v$ and diffusion coefficients $D$.
In order to include concentration profiles with different degrees of spatial dependence, quantified by $[F(0)-F(L)]/\fin $ (see Figure~\ref{fig1}C), several values of $D$ were chosen, and for each of them, a range of velocities $v$ was chosen using Figure~\ref{fig1}C so that it includes flat profiles for small velocities, spatial profiles for intermediate velocities, and again flat profiles close to the washout limit. Throughout, the food inflow $v\:\fin$ at the entrance of the gut was held constant to allow comparison. In a well-mixed system, we would have $\mathcal{F}=N_{\text{M}}/N_{\text{T}}$, where $N_{\text{M}}$ denotes the initial number of mutants in the system and $N_{\text{T}}$ the total number of bacteria in the system~\cite{Ewens79}. 
We find an excellent agreement with this expectation in the case of flat concentration profiles. This is evident for small $N_{\text{T}}$ values, which correspond to the largest velocities considered and thus to the washout limit, when the concentration profiles are the flattest (see Figure~\ref{fig3}). Conversely, in the strongly spatial regime (red symbols in Figure~\ref{fig3}), the fixation probability deviates from the well mixed system expectation, becoming substantially larger than it, and almost independent of the total population. For large $N_{\text{T}}$, which corresponds to small velocities, and hence flat profiles again, the fixation probability slowly converges back to the well-mixed system expectation (see Figure~\ref{fig3}). 

\newpage

\subsection*{The fixation probability of neutral mutants results from an active population}

\begin{figure}[h!]
\includegraphics[width=\textwidth]{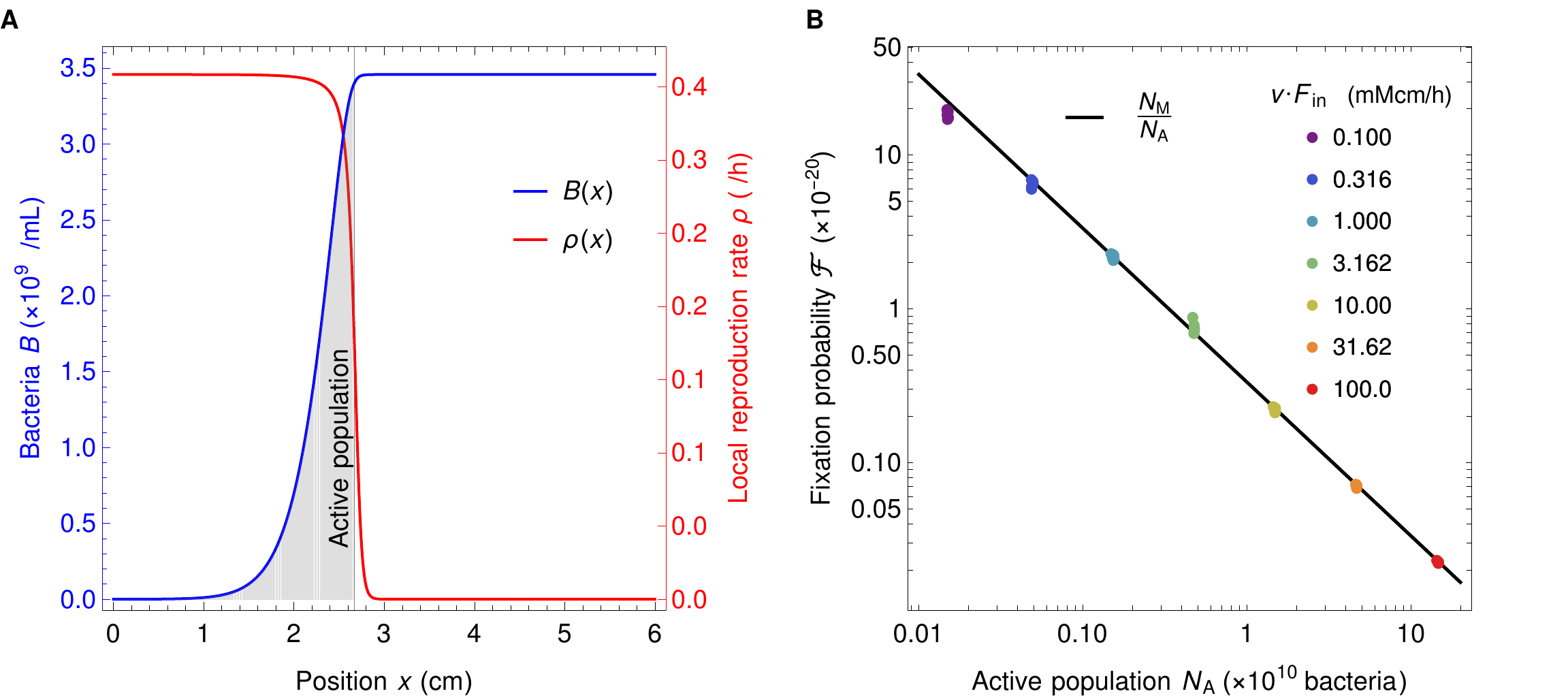}
\caption{\textbf{Active population explains the behavior of neutral mutant fixation probability in the gut.} \textbf{A}: The active bacterial population (gray shaded area) is defined as the total number of bacteria between the points $x=0$ and $x=x^*$, where $x^*$ is defined as $F(x^*)=k$, so that $B(x^*)=\alpha\fin (1-k/\fin)$. This corresponds to the region where bacteria have significant reproduction rates. Parameters are $v\:\fin=$ \SI[per-mode = symbol]{1}{\milli\Molar\centi\meter\per\hour}, $v=$ \SI[per-mode = symbol]{0.181}{\centi\meter\per\hour}, $D=$ \SI[per-mode = symbol]{0.02}{\centi\meter\squared\per\hour},  $k=$ \SI[per-mode = symbol]{0.1}{\milli\Molar}, $r=$ \SI{0.42}{\per\hour}, and $\alpha=$ \SI[per-mode = symbol]{0.613e9}{\bacteria\per\milli\Molar}.
\textbf{B}: Fixation probability $\mathcal{F}$ of neutral mutants in the gut in the regime with strong spatial dependence versus active population $N_{\text{A}}$, for different values of food inflow $v\:\fin$ (different colors). Each set of markers with a given color contains between 6 and 11 different points (often overlapping). 
Diffusion coefficient is the same for all points, $D=$ \SI[per-mode = symbol]{0.02}{\centi\meter\squared\per\hour}, velocities are
$v\in[0.135,0.171]$ \SI[per-mode = symbol]{}{\centi\meter\per\hour}, $v\in[0.14,0.18]$ \SI[per-mode = symbol]{}{\centi\meter\per\hour},
 $v\in[0.15,0.185]$ \SI[per-mode = symbol]{}{\centi\meter\per\hour}, $v\in[0.15,0.181]$ \SI[per-mode = symbol]{}{\centi\meter\per\hour}, 
 $v\in[0.15,0.186]$ \SI[per-mode = symbol]{}{\centi\meter\per\hour}, $v\in[0.155,0.182]$ \SI[per-mode = symbol]{}{\centi\meter\per\hour},
 $v\in[0.165,0.186]$ \SI[per-mode = symbol]{}{\centi\meter\per\hour}
for $v\:\fin=0.1$ to $100.0$ \SI[per-mode = symbol]{}{\milli\Molar\centi\meter\per\hour}, respectively. Other parameters and initial conditions are as in Figure~\ref{fig3}.
 Only the data points satisfying $\left[F(0)-F(L)\right]/\fin  >0.9$ are retained, ensuring that we focus on the plateau of the fixation probability with respect to the total population (see Fig.~\ref{fig3}). The black line corresponds to $\mathcal{F}= N_{\text{M}}/N_{\text{A}}$. Panel A corresponds to one of the green dots in panel B. }\label{fig4}
\end{figure}

Why is the fixation probability of neutral mutants larger in the gut in the presence of strong spatial dependence than in a well-mixed population with the same size? An important difference is that not all bacteria are actively reproducing in the gut, while they all have the same replication rate in a well-mixed population. More precisely, in the regime with strong spatial dependence, most replications occur in the region such that the local number of reproduction events $R(x)$ per unit volume and unit time (see Eqs.~\ref{eq:Reprrate3} and~\ref{eq:Reprrate2}) is substantial, i.e. visually, under the local replication rate curve, which coincides with the zone where bacterial concentration increases (see Figure~\ref{fig4}A).  Quantitatively, we define the ``active population'', i.e. the region with active reproduction, by comparing the replication rate to its maximum possible value, see Figure~\ref{fig4}A and the Supplementary material Section~\ref{SM:active_pop} for details. 

Can the active population, smaller than the total population, and comprising the reproducing bacteria, quantitatively explain the fixation probability observed in the gut in the presence of strong spatial dependence? In order to assess this, we set out to significantly change active population size, and thus the total number of reproduction events, by varying the food inflow $v\:\fin$ at the entrance of the gut, while holding the diffusion coefficient constant at $D=$\SI[per-mode=symbol]{0.02}{\centi\meter\per\hour}. We took several velocity values, but only retained those such that concentration profiles were strongly spatially dependent. Figure~\ref{fig4}B shows the fixation probability $\mathcal{F}$ versus the size $N_{\text{A}}$ of the active population in this spatial regime. Our results agree very well with the relation
\begin{equation}
	\mathcal{F} = \frac{N_{\text{M}}}{N_{\text{A}}}, \label{actfix}
\end{equation}
where $N_{\text{M}}$ is the initial number of mutant bacteria. This corresponds to the well-mixed expectation for the fixation probability of $N_{\text{M}}$ mutants in a population of $N_{\text{A}}$ bacteria, which confirms that the active population is the one that matters for the process of mutant fixation. This explains why the fixation probability is higher in the spatial system than in the well-mixed one, as well as the shape of the curves in Fig.~\ref{fig3}. Indeed, in each of these curves, as $v$ is decreased at a given $D$, $N_{\text{T}}$ increases. Fig.~\ref{fig1}C shows that the system then goes from quasi washout where $N_{\text{T}}$ is small and $N_{\text{A}}\approx N_{\text{T}}$ to the strongly spatial regime where $N_{\text{A}}\ll N_{\text{T}}$, and finally toward less strongly spatial regimes where replication is slow in the whole system. When $N_{\text{A}}\approx N_{\text{T}}$, Eq.~\ref{actfix} reduces to the well-mixed expectation $\mathcal{F}=N_{\text{M}}/N_{\text{T}}$, while it strongly deviates from it when $N_{\text{A}}\ll N_{\text{T}}$. Moreover, in the strongly spatial regime, almost all food is consumed. As the food inflow $v\,F_{in}$ is constant in Fig.~\ref{fig3}, the overall production rate of new bacteria, which is approximately $r N_{\text{A}}$ as most reproductions occur in the active population, is then constant too, yielding a constant $N_{\text{A}}$. Accordingly, Fig.~\ref{fig:Food_profiles}A shows that when $v$ is decreased within the strongly spatial regime, the transition from low to high bacterial concentration gradually shifts upstream in the gut while retaining the same shape, and thus $N_{\text{A}}$ remains constant while $N_{\text{T}}$ increases. This explains the plateau observed in Fig.~\ref{fig3}.

In the Supplementary material, section~\ref{SM:param}, we demonstrate the generality of the conclusions obtained here by systematically investigating the three dimensionless parameters that fully describe the stationary state of the system. Eq.~\ref{actfix} holds in all cases considered, provided that the food concentration profile is strongly spatial (see Figures~\ref{fig:NDPK},~\ref{fig:NDPR} and~\ref{fig:NDPL}). Furthermore, we demonstrate that the range of parameters considered in the present study matches the realistic one in the human colon. Finally, Fig.~\ref{fig:stoc_AP} in section~\ref{sec:stochsim} of the Supplementary material shows that our prediction in Eq.~\ref{actfix} is validated by stochastic simulations. 


\section*{Discussion}

We addressed bacterial evolution in the gut within a minimal model that incorporates flow and gradients of food and bacterial concentrations along the gut. We focused on the colon lumen, where the vast majority of our microbiota is located, and we studied parameter ranges relevant for the human colon. We considered neutral mutants appearing in the gut. Estimates of bacterial population sizes in the human colon~\citep{sender2016,claesson2009comparative} and of fitness effects~\citep{robert2018mutation} show that a substantial fraction of spontaneous mutations occurring in gut bacteria is expected to be effectively neutral (see Supplementary material, section~\ref{neutrality}). The dynamics of bacteria and food was described using a system of partial differential equations based on Refs.~\cite{Cremer2016,Cremer2017}. In the long term, in a finite-size system, mutants either disappear or take over due to stochastic fluctuations, and the stationary proportion of mutants in our continuous and deterministic description gives their fixation probability. We demonstrated that, in the regime where the profiles of food and bacterial concentrations are strongly spatial, with abundant food and few bacteria upstream, and vice-versa downstream, the stationary concentration of mutants is higher if they start upstream. However, for mutations occurring at replication, the small upstream concentration of bacteria means that few mutants appear there. Accordingly, we found that successful mutants are more likely to originate from an intermediate position along the gut. We studied the overall long term mutant proportion for neutral mutants appearing spontaneously upon division, which also gives their fixation probability. We found that in the  almost well-mixed regime, it is given by the ratio of the initial number of mutants to the total bacterial population size, consistently with the well-mixed expectation. By contrast, when the profiles of food and bacterial concentrations are strongly spatial, which is the relevant regime in the gut~\cite{Cremer2016,Cremer2017}, this fixation probability becomes substantially larger than the well-mixed expectation. Thus, the spatial structure of the gut favors the spread of neutral mutants and the evolution of the population composition. Furthermore, we rationalized this increase of the fixation probability by demonstrating that it stems from the fact that only a subset of the bacterial population is actively replicating. This active population is located upstream, where there is enough food to allow substantial replication. It gives an effective population size~\cite{Ewens79,Whitlock97} for the fixation of neutral mutants in the complex structured population of the gut.

Studies addressing the impact of spatial population structure on evolution generally consider discrete patches of population with migrations between them, and the same environment in each of them~\cite{Wright31,Kimura64,Maruyama70,Maruyama74,Nagylaki80,Slatkin81,Barton93,Whitlock97,Whitlock03,Lieberman05,Kaveh15, Hindersin15, Pattni15,Houchmandzadeh11,Houchmandzadeh13,Constable14,MarrecPreprint}. While complex population structures with asymmetric migrations can impact the fixation probabilities of beneficial and deleterious mutants~\cite{Whitlock03,Lieberman05,MarrecPreprint}, that of neutral mutants appearing uniformly in the population, e.g. upon division, is unaffected~\cite{Lieberman05,MarrecPreprint}. Similarly, chaotic hydrodynamic flow has been predicted to impact non-neutral mutant fixation probabilities, but not neutral ones~\cite{Herrerias18}. In the gut, the flow can be viewed as yielding asymmetric migrations.  Strikingly, we found that the fixation probability of neutral mutants could strongly differ from the well-mixed case. Aside the fact that the gut is a continuous system, a crucial difference with the above-cited models of population structure is that, due to directional hydrodynamic flow, the environment varies along the gut, in particular the food and bacterial concentrations, and thus the bacterial division rate. Environmental gradients can strongly impact evolution; for instance, gradients of antibiotics can increase the speed at which antibiotic resistance emerges \cite{Zhang11,Greulich2012, Hermsen12, Baym2016}. The coupling of bacterial concentration gradients due to antibiotics with convective flow also has complex implications on evolution~\citep{Gralka17}. Hydrodynamic flow itself can strongly impact fixation probabilities and fixation times, as has been shown in the case of compressible flows relevant for large-scale turbulent systems such as bacterial populations living at the surface of oceans~\citep{Perlekar10,Pigolotti12,Plummer19}. In these situations, flow reduces the effective population size for fixation probability, and microorganisms born near a flow source are more likely to fix than those born in a flow sink~\citep{Plummer19}. Albeit obtained in a different hydrodynamic regime, these results share similarities with ours, and together, they demonstrate that hydrodynamic flow, and in particular convective flow, can strongly impact evolution at various scales, from the gut to the ocean.

In addition to hydrodynamic flow and gradients, the gut comprises an upstream zone with few bacteria and rapid growth. This is reminiscent of expanding fronts in populations that invade a new environment~\citep{Hallatschek07,Hallatschek08}, which feature reduced competition and reduced effective population sizes, with important consequences on evolution~\citep{Travis07,Bosshard17,Gralka16}. In these cases, the dynamics is different depending on whether the traveling waves characterizing expansion are driven by the leading edge (pulled, as e.g. Fisher waves~\citep{Fisher37,KPP37,MurrayBook}) or by the bulk of the wave (pushed), yielding different wave velocities~\citep{Hallatschek08,Birzu18}. Contrary to population expansion on solid substrates~\citep{Hallatschek07}, the gut features directional hydrodynamic flow. The associated velocity $v$ is imposed, as opposed to a traveling wave velocity. Besides, boundary conditions put us in a different parameter regime compared to Fisher waves. One may worry that our deterministic model may not be appropriate in the upstream region. However, this concern is alleviated by the directional flow, which transports bacteria downstream. Specifically, bacteria take at least 20 minutes to replicate (here we took a typical replication time of 100 minutes), and since they are transported with the flow, the lineage of an upstream bacteria will be broadly distributed, including where there are many bacteria, before being large enough to affect $F$ sufficiently to modify the dynamical equations via $F/(k+F)$. Furthermore, the main findings from our deterministic model are validated by stochastic simulation results (Supplementary material section~\ref{sec:stochsim}).

Extending our study from neutral mutants to beneficial and deleterious ones, and studying fixation times and the rate of evolution in the gut, would be interesting topics for future work. Note that given the very large numbers of bacteria at play, fixation is expected to be slow. However, even before fixation, our results show that the proportion of mutants is increased by the gut structure compared to a well-mixed system. Our stochastic simulation results in Fig.~\ref{fig:stoc_times} confirm that the timescale for the increase of the average proportion of mutants is much faster than the one for mutant fixation.  Furthermore, while the minimal model used here captures some key characteristics of the gut, with a net flow, an effective mixing that is on limited length scales, and a stable bacterial population, the reality of the gut is more complex. In particular, muscle contractions in peristalsis and segmentation~\citep{Ailiani09,Huizinga09} mean that the radius of the gut is variable, and yield complex mixing dynamics. Besides, several food sources and several bacterial species are present, yielding complex ecological dynamics. Bacterial populations in the colon lumen can also interact with those in the mucus and in crypts. In addition, assuming a constant food inflow is a simplification, and in real life food inflow is variable, depending e.g. on the timing of meals, thus adding time variability to the spatial gradients we considered here. Despite all these complications, our results, which can be interpreted simply through the active population, have the potential to be general, and can be tested in more detailed models.

\paragraph{Note.}
While this manuscript was in revision, an independent and complementary study~\cite{ghosh2021emergent} was released on BioRxiv.

\paragraph{Acknowledgments.}
This work was supported by an Emergence Grant from Sorbonne Université (to A.-F. B.). A.-F. B. acknowledges funding by the European Research Council (ERC) under the European Union’s Horizon 2020 research and innovation programme (grant agreement No. 851173).


\bibliographystyle{unsrt}

\newpage

{\LARGE \bf Supplementary material}

\tableofcontents

\renewcommand{\thesection}{S\arabic{section}}

\renewcommand{\thefigure}{S\arabic{figure}}
\setcounter{figure}{0}
\renewcommand{\thetable}{S\arabic{table}}
\setcounter{table}{0}

\section{Numerical methods}\label{SM:num_meth}

\subsection{Discretization of the partial differential equation system}\label{SM:dicretization}

In order to solve the system~\ref{eq:PDEs} numerically, we employ an explicit method using a forward difference for the time derivative at time $t$ and a central difference for the space derivative at position $x$. Explicitly, the relevant differential operators are replaced by the following expressions:
\begin{subequations}\label{eq:discret}
\begin{align}
	\frac{\partial Y}{\partial t} & \rightarrow  \frac{Y(t,x)-Y(t-\Delta t,x)}{\Delta t},  \\
	\frac{\partial Y}{\partial x} & \rightarrow  \frac{Y(t,x+\Delta x)-Y(t,x-\Delta x)}{2\Delta x},  \\
	\frac{\partial^2 Y}{\partial x^2} & \rightarrow  \frac{Y(t,x+\Delta x)-2Y(t,x)+Y(t,x-\Delta x)}{(\Delta x)^2},  
\end{align} 
\end{subequations}
where $\Delta t$ and $\Delta x$ represent the discrete steps in time and space, respectively. Here $Y(t,x)$ can represent the concentration of food $F$, or  wild type bacteria $B$, or mutant bacteria $M$, at time $t$ and coordinate $x$. Substituting the differential operators in Eqs.~\ref{eq:PDEs} using Eqs.~\ref{eq:discret} yields
\begin{subequations}\label{eq:sys_discr}
\begin{align}
	F(t+\Delta t,x) & =  F( t,x) +  D \frac{F(t,x+\Delta x)-2F(t,x)+F(t,x-\Delta x)}{(\Delta x)^2}\Delta t -\nonumber \\
	& -  v \frac{F(t,x+\Delta x)-F(t,x-\Delta x)}{2\Delta x}\Delta t -\frac{r}{\alpha}(B(t,x)+M(t,x) )\frac{F(t,x)}{k+F(t,x)}\Delta t,  \\
	B(t+\Delta t,x) & =  B( t,x) +  D \frac{B(t,x+\Delta x)-2B(t,x)+B(t,x-\Delta x)}{(\Delta x)^2}\Delta t-\nonumber \\
	& - v \frac{B(t,x+\Delta x)-B(t,x-\Delta x)}{2\Delta x}\Delta t +r B(t,x)\frac{F(t,x)}{k+F(t,x)}\Delta t,  \\
	M(t+\Delta t,x) & =  M( t,x) +  D \frac{M(t,x+\Delta x)-2M(t,x)+M(t,x-\Delta x)}{(\Delta x)^2}\Delta t - \nonumber \\
	& v \frac{M(t,x+\Delta x)-M(t,x-\Delta x)}{2\Delta x}\Delta t +r M(t,x)\frac{F(t,x)}{k+F(t,x)}\Delta t.
\end{align}
\end{subequations}
The boundary conditions in $x=0$ from Eqs.~\ref{eq:PDEs_BC} become
\begin{subequations}
\begin{align}
	F(t,0-\Delta x) &= F(t,0+\Delta x) + \frac{2\Delta x}{D}v\left[ \fin   - F(t,0) \right], \label{eq:PDEs_BC_A} \\
	B(t,0-\Delta x) &= B(t,0+\Delta x) -  \frac{2\Delta x}{D}v B(t,0),  \\
	M(t,0-\Delta x) &= M(t,0+\Delta x) -  \frac{2\Delta x}{D}v M(t,0), 
\end{align}\label{eq:sys_discr_LB}
\end{subequations}
while the boundary conditions in $x=L$ from Eqs.~\ref{eq:PDEs_BC} become
\begin{subequations}
\begin{align}
	F(t,L+\Delta x) &= F(t, L-\Delta x),\\
	B(t,L+\Delta x) &= B(t, L-\Delta x), \\
	M(t,L+\Delta x) &= M(t, L-\Delta x). 
\end{align}\label{eq:sys_discr_RB}
\end{subequations}

The spatial discrete step $\Delta x$ is in general chosen to be \SI{0.01}{\centi\meter}. This value is small enough to ensure convergence for most model parameters, and to have a good spatial resolution for analysis. Note however that for some parameters,  the term $\left[ \fin - F(t,0) \right]$ in  Eq.~\ref{eq:PDEs_BC_A} can be very large which can lead to numerical instability. To compensate, the spatial step, $\Delta x$, needs to be reduced. 
Once $\Delta x$ is chosen, the value of $\Delta t$ should satisfy the stability condition $\Delta t \leq (\Delta x)^2/(2 D)$ \cite{crank1979mathematics}. Specifically, we take $\Delta t =0.8 (\Delta x)^2/(2 D)$.  

\subsection{Obtaining stationary profiles without and with mutants}\label{SM:stat_prof_num}

Numerically, we determine the unique solution for food concentration that satisfies $F(x)<\fin$ for all $x$. Thus, the steady-state profile of food concentration is independent of the initial conditions as long as they are all positive. In general, for our numerical integration, we choose initial conditions not too far to the steady state, namely $F(0,x)=F_0\in(0,\fin)$, $B(0,x)=\alpha[\fin-F(0,x)]$, and $M(0,x)\ll B(0,x)$, in order to obtain faster convergence. If there is no mutant bacteria, the steady state of the wild type bacteria concentration is also uniquely defined through the relation $B(x)=\alpha [\fin-F(x)]$. However, if there are both wild type and mutant bacteria in the system, then the steady state solution is uniquely defined only for the total bacterial concentration, while individual concentrations depend on the initial conditions for wild type and mutant bacteria. 

In practice, the stationary state in the absence of mutant bacteria is found using Eqs.~\ref{eq:sys_discr} coded in the Fortran90 programming language (code available at \texttt{https://doi.org/10.5281/zenodo.4704653}~\cite{codeZenodo}) with homogeneous initial conditions
\begin{subequations}
     \begin{align}
        F(0,x) & = 0.9\fin,         \\
        B(0,x) & = \alpha[\fin-F(0,x)] = 0.1 \alpha\fin        . 
     \end{align}
\end{subequations}
The time used in all numerical integrations is $t=$ \SI{500}{\hour}, which is long enough for the system to reach the steady state for all choices of parameters considered in this paper. 

To find the mutant concentration profile that is crucial to our calculation of the fixation probability, we consider the system without mutants at steady state, and we assume that mutants appear at one local position, $\xM$, in the gut segment. We again solve Eqs.~\ref{eq:sys_discr} for time $t=$ \SI{500}{\hour}, but now with the initial conditions
\begin{subequations}
     \begin{align}
         F(0,x) &= F^*(0,x), \\
         B(0,x) &= B^*(0,x), \\
         M(0,x) &=\begin{cases} M_0, \quad \left|x-\xM\right|\leq \Delta x/2,\\ 0,\quad\quad \left|x-\xM\right|>\Delta x/2, \end{cases}
     \end{align}
\end{subequations}
where $F^*$ and $B^*$ represent steady state concentrations without mutant bacteria, while $M_0\ll B(x_M)$  is the initial local mutant concentration. More precisely, denoting by $N_{\text{M}}$ the total number of mutant bacteria introduced in the system, the quantity $M_0$ is the initial concentration in the segment $[\xM-\Delta x/2,\xM+\Delta x/2]$, where $\Delta x$ is the spatial discrete step of our numerical resolutions.
Hence, $M_0$ satisfies $N_{\text{M}} M_0 S\Delta x$, where $N_{\text{M}}$ is the total number of mutants introduced in the system. Note that since we are using central difference discretization, we need to double the $M_0$ value on boundaries of the segment where the mutants are introduced, in order to have the same $N_{\text{M}}$ there as in the rest of the segment.
In practice, we choose the value $N_{\text{M}} =3.33\times 10^{-11}$ bacteria, so that for any choice of parameters used in this paper and for any initial mutant position $\xM$, the relation $M_0\ll B(\xM)$ is satisfied. Importantly, since the stationary concentration of mutant bacteria is proportional to $M_0$, all results scale with it, and we are not losing generality by fixing the value of $M_0$.

\subsection{Conversion between food and bacteria concentrations}\label{SM:OD_conversion}

The initial unit of food concentration is moles per liter, and that of bacterial concentration is optical density, \SI{}{\OD}~\cite{Cremer2016}, which can be converted to numbers of bacteria per volume by using the calibration curve in~\cite{Cremer2016}. Specifically, the conversion factor we take is  \SI{1}{\OD}=\SI[per-mode = symbol]{3.33e9}{\bacteria\per\milli\liter}. Then the parameter $\alpha$ allows to convert between food and bacteria concentrations.

Importantly, because $\alpha$ is just a scaling factor, a change in this value will modify the bacterial concentration quantitatively, but the spatial profile and all the other conclusions will remain identical.

\section{Stationary profiles without mutants}\label{SM:stat_prof}

\subsection{Ordinary differential equation description} \label{SM:ODEs}

Without mutants, at stationary state, Eq.~\ref{eq:PDEs} yields:
\begin{align}
0&=D\frac{\partial^2 F}{\partial x^2}-v\frac{\partial F}{\partial x}-\frac{r}{\alpha}\frac{FB}{k+F}\,,\\
0&=D\frac{\partial^2 B}{\partial x^2}-v\frac{\partial B}{\partial x}+r\frac{FB}{k+F}\,,
\end{align}
with boundary conditions in Eq.~\ref{eq:PDEs_BC}.

Introducing $f=\alpha F+B$, we have at stationary state
\begin{equation}
C=D\frac{\partial f}{\partial x}-vf\,,
\end{equation}
where $C$ is a constant. The solution reads
\begin{equation}
f(x)=-\frac{C}{v}+C' e^{vx/D}\,,
\end{equation}
where $C'$ is a constant. Applying the boundary conditions yields $C'=0$ and $C=-v\alpha \fin $. Thus,
\begin{equation}
\alpha F(x)+B(x)=f(x)=\alpha \fin\,,
\end{equation}
and this specific linear combination of $B$ and $F$ is independent from $x$: 
food effectively gets converted into bacteria.

Now we can inject this into the equation on $F$ to decouple it from $B$, yielding:
\begin{equation}
0=D\frac{\partial^2 F}{\partial x^2}-v\frac{\partial F}{\partial x}-r\frac{F(\fin -F)}{k+F}\,,
\label{edo}
\end{equation}
which is a second order nonlinear ordinary differential equation.

\subsection{Dimensionless form}\label{SM:dimless}

Let us make the variable change $s=xv/D$, and let us introduce the function $\phi$ satisfying  $\phi(s=xv/D)=F(x)/\fin$ for all $x$. We obtain
\begin{equation}
0=\frac{\partial^2 \phi}{\partial s^2}-\frac{\partial \phi}{\partial s}-\lambda\frac{\phi(1-\phi)}{\kappa+\phi}\,,
\label{simplifiedEq}
\end{equation}
which involves the dimensionless numbers
\begin{equation}
\kappa=\frac{k}{\fin}\,,
\end{equation}
and 
\begin{equation}
\lambda=\frac{rD}{v^2}\,. 
\end{equation}
The associated boundary conditions are:
\begin{align}
\phi(s=0)-\frac{\partial \phi}{\partial s}(s=0)&=1\,,\\
\frac{\partial \phi}{\partial s}\left(s=\sigma\right)&=0\,
\end{align}
where 
\begin{equation}
\sigma=\frac{Lv}{D}
\end{equation} 
is the third dimensionless number describing the system~\cite{Cremer2016}, and compares transport by convection to transport by diffusion. It has the same form as the Péclet number for mass transfer, but recall that $D$ is an effective diffusion coefficient modeling gut contractions.

\subsection{Some stationary profiles}\label{SM:stat_prof_ex}

\begin{figure}[h!]
\includegraphics[width=\textwidth]{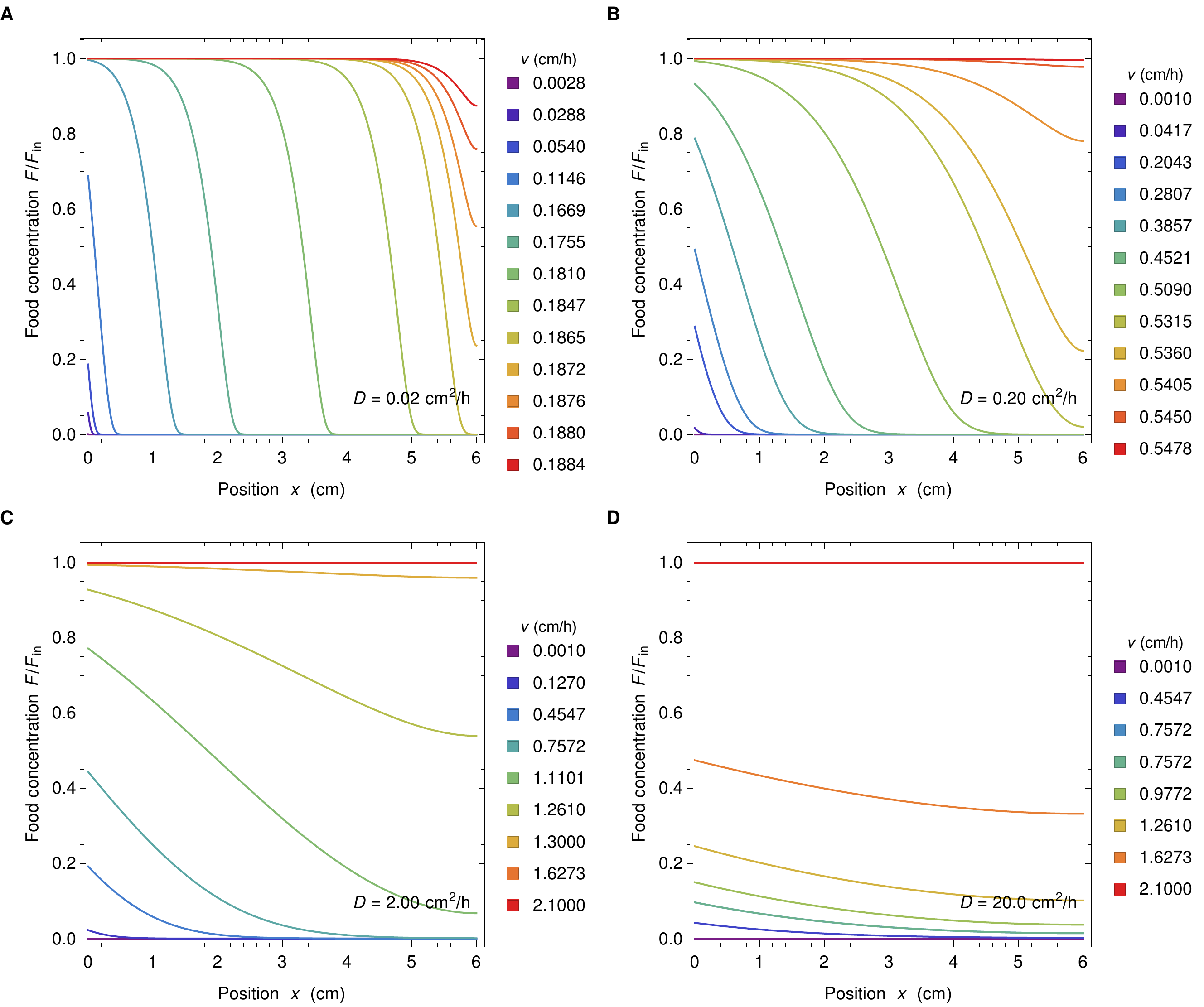}
\caption{
\textbf{Food concentration profiles.} Food concentration $F$ normalized by the food concentration inflow $\fin$ versus position $x$ along the gut for four different values of the diffusion coefficient (panels A to D) and several values of velocity (different colors in each panel). Values of diffusion coefficients and velocity are indicated in each panel. Other parameter values are $k=$ \SI[per-mode = symbol]{0.1}{\milli\Molar}, $r=$ \SI{0.42}{\per\hour}, $v\:\fin=$ \SI[per-mode = symbol]{1}{\milli\Molar\centi\meter\per\hour}, $\alpha=$ \SI[per-mode = symbol]{6.13e8}{\bacteria\per\milli\Molar}.  }\label{fig:Food_profiles}
\end{figure}

In practice, the partial differential equations in Eqs.~\ref{eq:PDEs} with boundary conditions in Eqs.~\ref{eq:PDEs_BC} and initial conditions in Eqs.~\ref{eq:PDEs_IC} were solved numerically as explained above. Examples of profiles obtained are given in Figure~\ref{fig:Food_profiles}. Note that we checked that the long-term results from their direct resolution was consistent to those obtained by numerically solving the ordinary differential equation \ref{edo} giving the stationary state of the system.

\clearpage

\section{Comparison to Fisher waves}
\label{secFisher}

In the regime with strong spatial dependence, the present model yields at steady state an upstream zone with few bacteria, high food concentration and rapid growth, and a downstream zone with many bacteria, little food and little growth. Specifically, our numerical resolutions show that $\phi=F/\fin$ then monotonically decays from a value close to one to a value close to zero, with a sigmoid-like shape (see Fig.~\ref{fig1}B, as well as Fig.~\ref{fig:Food_profiles}A-B for intermediate velocities). This is reminiscent of Fisher waves~\citep{Fisher37}, which are (among several applications) an important model for the spread of mutant genes in populations~\citep{Fisher37}, as well as for the expansion of populations that invade a new environment~\citep{FifeBook,MurrayBook}, where the dynamics of mutants has been extensively discussed~\citep{Hallatschek07,Hallatschek08,Travis07,Korolev10,Bosshard17,Gralka16,Birzu18}. Let us compare these two models.

The Fisher-Kolmogorov-Petrovsky-Piskunov (Fisher-KPP) equation~\citep{Fisher37,KPP37} is a reaction-diffusion equation of the form~\citep{FifeBook,PerthameBook}
\begin{equation}
\frac{\partial u}{\partial t}=\frac{\partial^2 u}{\partial x^2}+f(u)\,,
\label{fkpp}
\end{equation}
where $f$ is a function satisfying $f(0)=f(1)=0$.
For $f(u)=u(1-u)$, it is known as the Fisher equation~\citep{Fisher37,MurrayBook}.
Searching for a ``Fisher wave'' solution of Eq.~\ref{fkpp} in the form of a traveling wave $u(x,t)=U(z=x\pm ct)$, where $c$ is a wave velocity to be determined~\citep{FifeBook}, yields:
\begin{equation}
\frac{d^2 U}{d z^2}\pm c\frac{d U}{d z}+f(U)=0\,.
\label{fkpp2}
\end{equation}
It is known~\cite{PerthameBook} that if $f'(0) > 0$, $f'(1) < 0$, and $f(x)>0$ for all $x\in\,]0,1[$, then monotonic wave front solutions such that $\lim_{z\to-\infty}U(z)=0$ or 1, and $\lim_{z\to\infty}U(z)=1$ or 0, exist if and only if $|c|\geq c^*$, where $c^*>0$ is a constant that can be determined by analyzing the stability of the fixed points at $(U,U')=(0,0)$ and $(U,U')=(1,0)$. In particular, $c^*=2$ in the case of the Fisher equation~\citep{MurrayBook}.

In our model, steady-state profiles of $\phi$ are solutions of Eq.~\ref{simplifiedEq}, and thus, using $\xi=x/D$ instead of $s=xv/D$, the function $\psi=1-\phi$ satisfies
\begin{equation}
\frac{d^2 \psi}{d \xi^2}-v\frac{d \psi}{d \xi}+rD\frac{\psi(1-\psi)}{\kappa+1-\psi}=0\,,
\label{simplifiedEqter}
\end{equation}
which has the form of Eq.~\ref{fkpp2}, with $v$ replacing $\mp c$, and with $f:\psi\mapsto rD\,\psi(1-\psi)/(\kappa+1-\psi)$. This function satisfies $f(0)=f(1)=0$, as well as $f'(0) > 0$, $f'(1) < 0$, and $f(x)>0$ for all $x\in\,]0,1[$, since $r$, $D$ and $\kappa$ are positive constants. Therefore, monotonic stationary solutions such that $\lim_{z\to-\infty}\psi(z)=0$ or 1, and $\lim_{z\to\infty}\psi(z)=1$ or 0, exist if and only if $v\geq v^*$ (recall that in our system $v>0$), and the same result holds for $\phi=1-\psi$. 

\begin{figure}[h!]
\includegraphics[width=\textwidth]{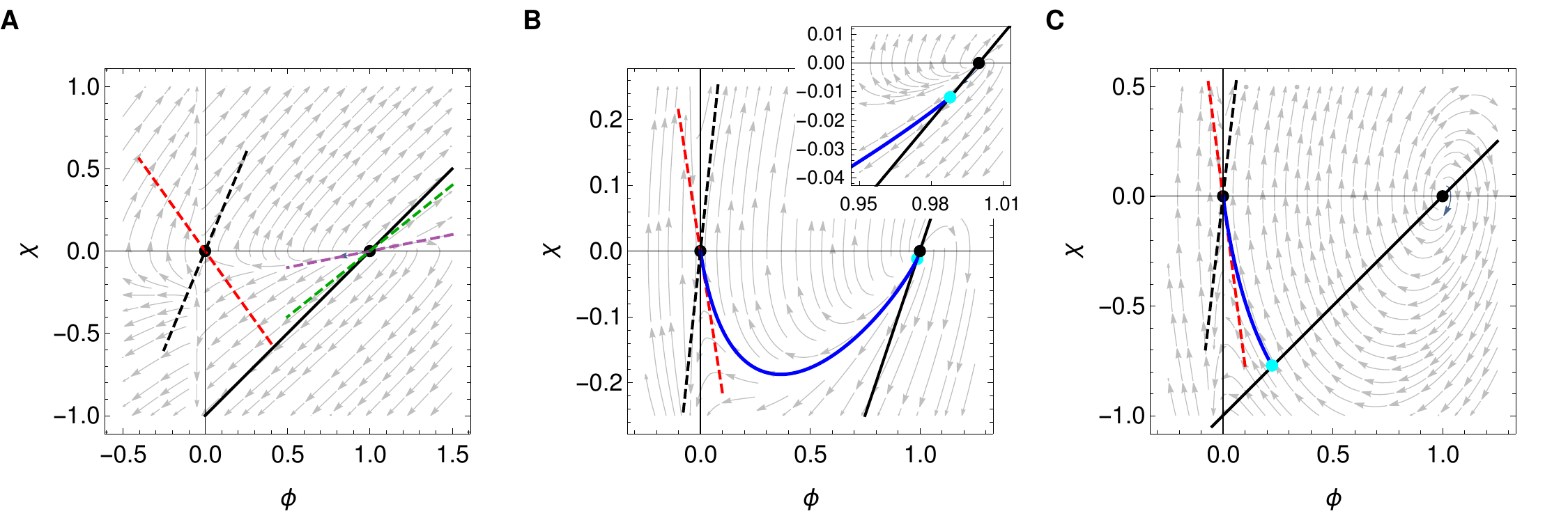}
\caption{
\textbf{Phase space analysis of the system.} \textbf{A}: $\lambda=0.17$ and $\kappa=0.05$, such that $v>v^*$ (washout). \textbf{B}: $\lambda=0.34$ and $\kappa=0.05$, such that $v<v^*$ (below washout, but close to it), corresponding to the parameters of Fig.~\ref{fig1}B.  \textbf{C}: $\lambda=3.36$ and $\kappa=0.05$, such that $v<v^*$ (substantially below washout). In all panels: Arrows: stationary solutions. Blue curve: stationary solution obtained by numerical integration of PDE Eqs.~\ref{eq:PDEs}.
 Black markers: fixed points. Dashed lines: directions of the Jacobian eigenvectors at these fixed points (at $(0,0)$: stable direction in red, unstable one in black; at $(1,0)$ if it is an unstable node (i.e. in \textbf{A}): slow unstable direction in purple, fast one in green). Black line: boundary condition in $s=0$, namely $\chi=\phi-1$. Cyan dot: Crossing of the numerical solution (blue curve) and the boundary condition (black line). Parameters of the numerical solutions are $D=0.1,\,0.2,\,2.0$~\SI[per-mode = symbol]{}{\centi\meter\squared\per\hour} in panels \textbf{A} , \textbf{B}, and \textbf{C}, respectively. Other parameters are as in Fig.~\ref{fig1}B. The inset in panel \textbf{B} shows a close-up in the vicinity of $(1,0)$.}\label{fig:phase}
\end{figure}

In order to determine $v^*$, let us study the stability of the fixed points at $(\phi,\phi')=(0,0)$ and $(\phi,\phi')=(1,0)$ in Eq.~\ref{simplifiedEq}. Introducing $\chi=\phi'$, where the prime denotes a derivative with respect to $s=xv/D$, Eq.~\ref{simplifiedEq} becomes:
\begin{equation}
    \begin{cases}
      \phi'=\chi\,,\\
      \chi'=\chi+\lambda\, \phi(1-\phi)/(\kappa+\phi)\,.
    \end{cases} 
\end{equation}
A linear stability analysis demonstrates that $(0,0)$ is a saddle point, while $(1,0)$ is either an unstable node if $4\lambda<\kappa+1$ (see Fig.~\ref{fig:phase}A) or an unstable spiral if $4\lambda>\kappa+1$ (see Fig.~\ref{fig:phase}B and C). Qualitative phase space analysis (similar to that presented e.g. in~\cite{MurrayBook} for the Fisher equation) then shows that stationary solutions such that $\lim_{s\to-\infty}\phi(s)=1$ and $\lim_{s\to\infty}\phi(s)=0$ exist in both cases, but that they are monotonic if and only if $4\lambda\leq\kappa+1$, or equivalently
\begin{equation}
v\geq v^*=\sqrt{\frac{4rD}{\kappa+1}}\,.
\label{vstar}
\end{equation}
Therefore, in infinite space, there exists a stationary solution to the system of partial differential equations Eqs.~\ref{eq:PDEs} that is equivalent to a Fisher wave (in a moving frame), provided that Eq.~\ref{vstar} is satisfied. However, the velocity $v$ is a parameter of the system, and not a wave velocity that can adjust, in contrast to the usual Fisher-KPP case. 

The model studied in this paper employs the system of partial differential equations Eq.~\ref{eq:PDEs} in a finite segment $x\in[0,L]$ (or equivalently, $s\in[0,\sigma=Lv/D]$), with the boundary conditions Eq.~\ref{eq:PDEs_BC}, which become in dimensionless form:
\begin{equation}
    \begin{cases}
      \chi(0)=\phi(0)-1\,,\\
      \chi(\sigma)=0\,.
    \end{cases} 
    \label{bcn}
\end{equation}
Is there a nontrivial solution satisfying these conditions, and if yes, how does it compare to the Fisher-wave-like monotonic solution such that $\lim_{s\to-\infty}\phi(s)=1$ and $\lim_{s\to\infty}\phi(s)=0$? 

First assume that Eq.~\ref{vstar} is satisfied, ensuring that the Fisher-wave-like solution exists (see Fig.~\ref{fig:phase}A). Then, $(1,0)$ is an unstable node and the smallest eigenvalue of the associated Jacobian is $\ell_-=[1-\sqrt{1-4\lambda/(\kappa+1)}]/2$. This eigenvalue satisfies $0\leq\ell_-\leq1/2$, and is associated to the eigenvector $(1,\ell_-)$, yielding a slow direction in the phase space with slope between 0 and $1/2$. Close to $(1,0)$, the Fisher-wave-like solution follows this slow direction, before curving upwards to finally approach the stable direction of the saddle point in $(0,0)$. Therefore, it only crosses the line $\chi=\phi-1$ corresponding to the boundary condition in $s=0$ (see Eq.~\ref{bcn}) at the unstable node $(1,0)$. But this unstable node can only be approached at $s\to-\infty$ in nontrivial solutions, which makes it impossible to find a nontrivial solution satisfying Eq.~\ref{bcn}. And indeed, our numerical resolutions confirm that if Eq.~\ref{vstar} is satisfied, bacteria are washed out (see Fig.~\ref{fig1}C and section~\ref{SM:washout}).

\begin{figure}[h!]
\includegraphics[width=\textwidth]{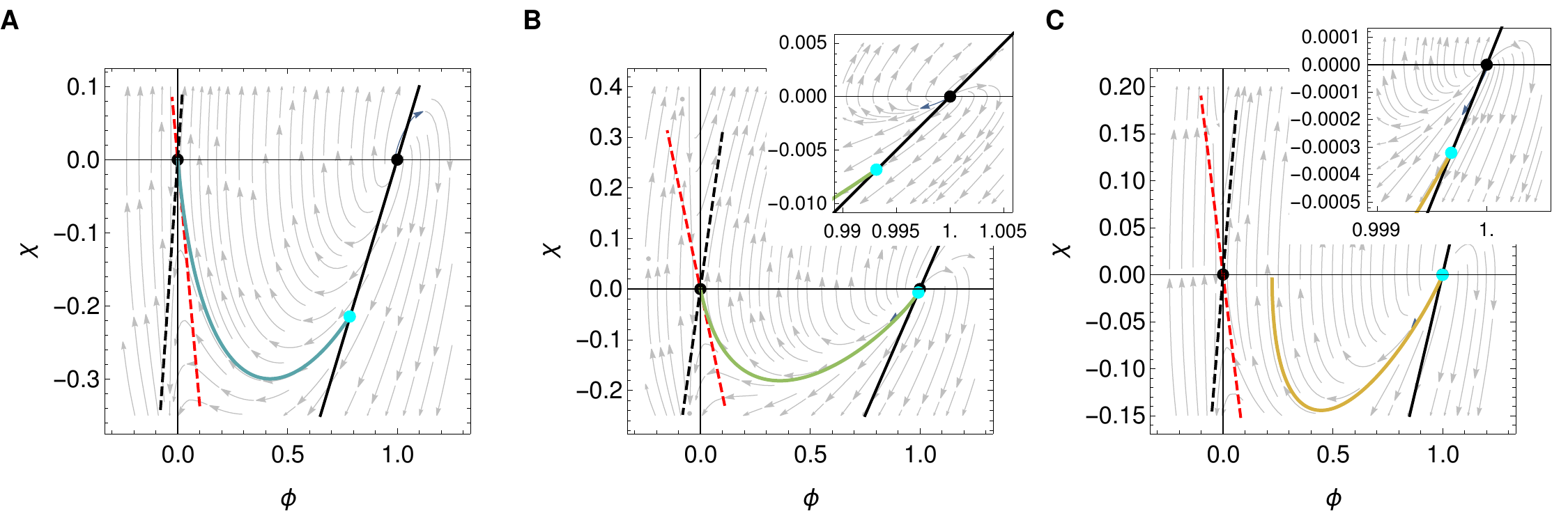}
\caption{
\textbf{Phase portraits for different types of spatial profiles from Fig.~\ref{fig:Food_profiles}B.}
\textbf{A:} $\lambda=0.56$ and $\kappa=0.039$.
\textbf{B:} $\lambda=0.32$ and $\kappa=0.051$.
\textbf{C:} $\lambda=0.29$ and $\kappa=0.054$.
Parameters of the numerical solutions are $v=0.386,\,0.509,\,0.536$~\SI[per-mode = symbol]{}{\centi\meter\per\hour} in panels \textbf{A} , \textbf{B}, and \textbf{C}, respectively. Other parameters are as in Fig.~\ref{fig1}B. Symbols are the same as in Fig.~\ref{fig:phase} and colors of curves representing numerical solutions match those of Fig.~\ref{fig:Food_profiles}B. Insets in panels \textbf{B} and \textbf{C} are close-ups in the vicinity of $(1,0)$.
}\label{fig:phase-v}
\end{figure}

\begin{figure}[h!]
\includegraphics[width=.9\textwidth]{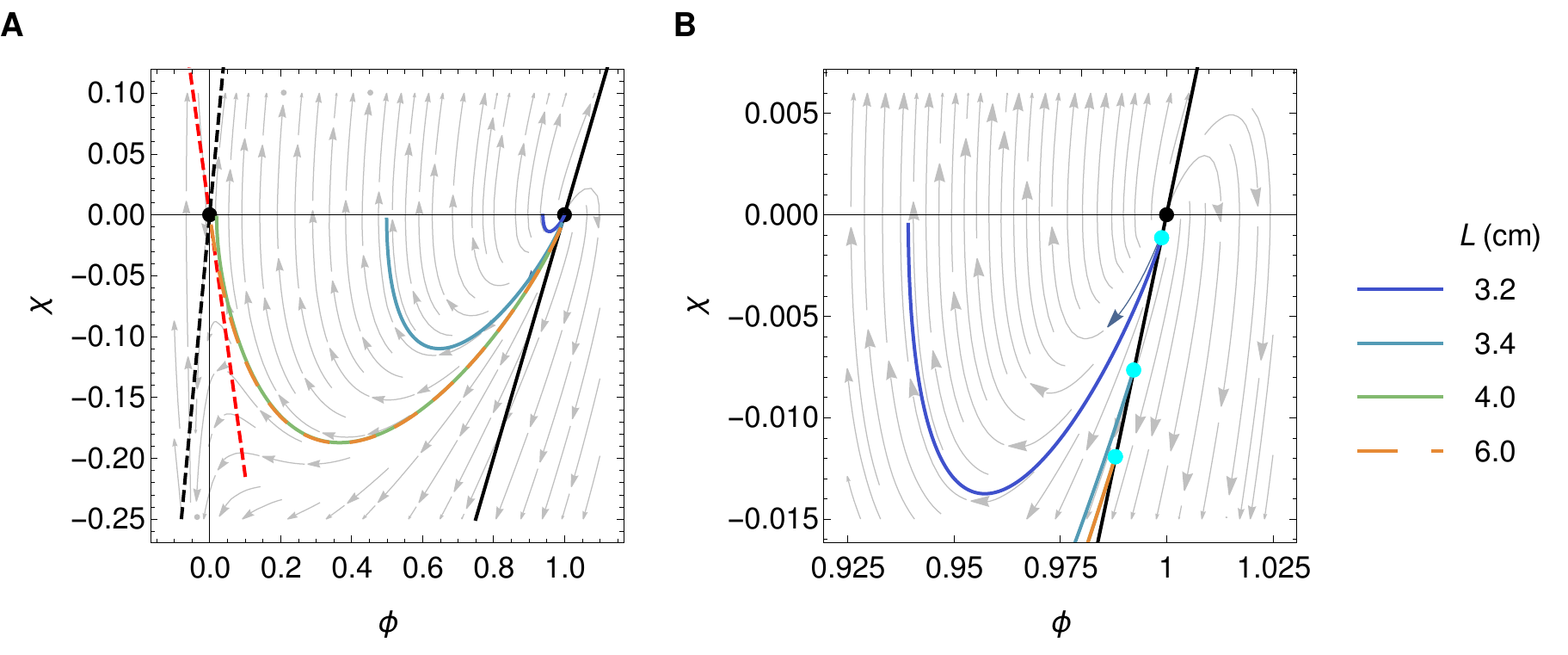}
\caption{
\textbf{Phase portraits for different system lengths $L$.}
All other parameters are as in Fig.~\ref{fig1}B. Symbols are the same as in Fig.~\ref{fig:phase} and colors of curves representing numerical solutions match those of Fig.~\ref{fig:NPF}B. Panel {\bf B} is a close-up of panel {\bf A} in the vicinity of $(1,0)$.
}\label{fig:phase-L}
\end{figure}

Let us turn to the opposite case, where $v<v^*$ (see Fig.~\ref{fig:phase}B and C). Then, $(1,0)$ is an unstable spiral, and thus the stationary solution such that $\lim_{s\to-\infty}\phi(s)=1$ and $\lim_{s\to\infty}\phi(s)=0$ is oscillatory and comprises unphysical parts where $\phi>1$ (implying negative numbers of bacteria). However, this spiraling behavior means that this special solution crosses the line $\chi=\phi-1$ (see Eq.~\ref{bcn}) in at least another point, say $P$, than the unstable node $(1,0)$. Thus, there may be nontrivial and physical solutions satisfying Eq.~\ref{bcn}, among the solutions that intersect the line $\chi=\phi-1$ between $(1,0)$ and $P$ (examples of these intersections are shown in cyan in Figs.~\ref{fig:phase},~\ref{fig:phase-v} and~\ref{fig:phase-L}). These solutions may look quite different from the Fisher-wave like solution that exists for $v\geq v^*$. In particular, they may possess a value of $\phi(0)$ substantially smaller than 1 and a value of $\phi(\sigma)$ substantially larger than 0. These differences are expected to increase as $v$ is decreased further and further from $v^*$ (i.e. more differences from the Fisher-wave-like solutions are expected in Fig.~\ref{fig:phase}B than in Fig.~\ref{fig:phase}C). Consistently, Fig.~\ref{fig1}C demonstrates that profiles with strongest spatial dependence are obtained for $v$ not much below the washout limit $v=v^*$, and Fig.~\ref{fig:Food_profiles} includes several numerical solutions with values substantially different from 1 and 0 at the boundaries $s=0$ and $s=\sigma$. Fig.~\ref{fig:phase-v} shows the phase portraits corresponding to some of these cases, illustrating the diversity of solutions. Nevertheless, in the strongly spatial regime where $v$ is not much below the washout limit $v=v^*$, similarities with the Fisher-wave-like solutions that exist for $v>v^*$ are expected. This case is illustrated e.g. by Fig.~\ref{fig:phase-v}B.

Here, differences with Fisher waves are brought by the boundary conditions, and in particular by the one at $x=0$. In addition to the different parameter regime where nontrivial solutions exist, another difference is that all horizontal translations of a traveling wave solution are also solutions in the Fisher wave case~\cite{MurrayBook}, while here, they would not satisfy the boundary conditions. The boundary condition at $x=0$ is quite important as it corresponds to the transition between the small intestine and the colon. Besides, the boundary condition in $x=L$ also affects the solution, in particular by inducing a second washout limit, see Fig.~\ref{fig1}C and section~\ref{SM:washout}. However, for sufficiently large $\sigma=Lv/D$, the value of $\sigma$ has little impact on the solution (see Fig.~\ref{fig:NDPL}). Fig.~\ref{fig:phase-L} shows phase portraits with different values of $L$, illustrating its impact for small $L$, and its lack of impact for $L\geq 4$~cm. 

Note that traveling waves arising in Fisher-KPP equations Eq.~\ref{fkpp} (including those characterizing expanding populations) can be either driven by the leading edge (pulled) or by the bulk of the wave (pushed)~\citep{Hallatschek08,Birzu18}. Pulled waves have a velocity that depends only on $f(0)$, and are obtained when $f'(u)\leq f'(0)$ for all $u\in[0,1]$ and $f'(u)= f'(0)+O(h^p)$ for some $p>0$ when $u\to 0$, in addition to the above conditions $f'(0) > 0$, $f'(1) < 0$, and $f(x)>0$ for all $x\in\,]0,1[$~\cite{Bramson83,BrunetHDR}. These conditions are all satisfied by $f$ in Eq.~\ref{simplifiedEqter}, and thus the associated Fisher-KPP equation would lead to pulled waves. Pushed waves have a velocity that depends on the full nonlinearity of $f$, and can be obtained when growth rates are nonmonotonic with population density, which can occur with cooperativity (Allee effect)~\citep{Hallatschek08,Birzu18}. However, here, the velocity $v$ is an imposed parameter, in contrast to a traveling wave velocity, and the boundary conditions yield further differences, as discussed above. 

Note also that it would be interesting to further investigate the stability of solutions to time-dependent perturbations, as was done e.g. for the Fisher wave in the Fisher equation~\cite{MurrayBook}. Here, the analysis would be different as one would need to start from the system of partial differential equations Eqs.~\ref{eq:PDEs}.

\section{Washout limits}\label{SM:washout}

The washout limits are the limits where all bacteria get washed out of the system. Mathematically, they correspond to a bifurcation point in the parameter space where the trivial steady state solution $F(x)= \fin$ (and $B(x)=0$) becomes stable. 

In the chemostat, this limit is easy to determine. There are two steady state solutions, $F^\ch=kc/(r-c)$ and $F^*=\finc$. Eigenvalues of the Jacobian associated to Eqs.~\ref{chemo1} and~\ref{chemo2} for both steady states change their sign at same point in the parameter space, namely at
\begin{equation}
 c = \frac{r}{k/\finc+1}\equiv c_{\text{wo}}.
\end{equation} 
The bifurcation scenario is such that at $c_{\text{wo}}$ the two steady state solutions collide, change stability, and one (nontrivial) solution disappears, making it a transcritical bifurcation.

In the spatial system, a complete analytical treatment is more difficult, even though one of the washout limits was found by our phase portrait analysis above (see Eq.~\ref{vstar}). Numerically, we find one positive nontrivial solution for $v<v_{\text{wo}}$, where $v_{\text{wo}}$ is the bifurcation point which depends on the rest of the system parameters, and only the trivial solution for  $v>v_{\text{wo}}$, implying a change of the stability at $v=v_{\text{wo}}$ when the two steady states collide.

\begin{figure}[h!]
\includegraphics[width=.7\textwidth]{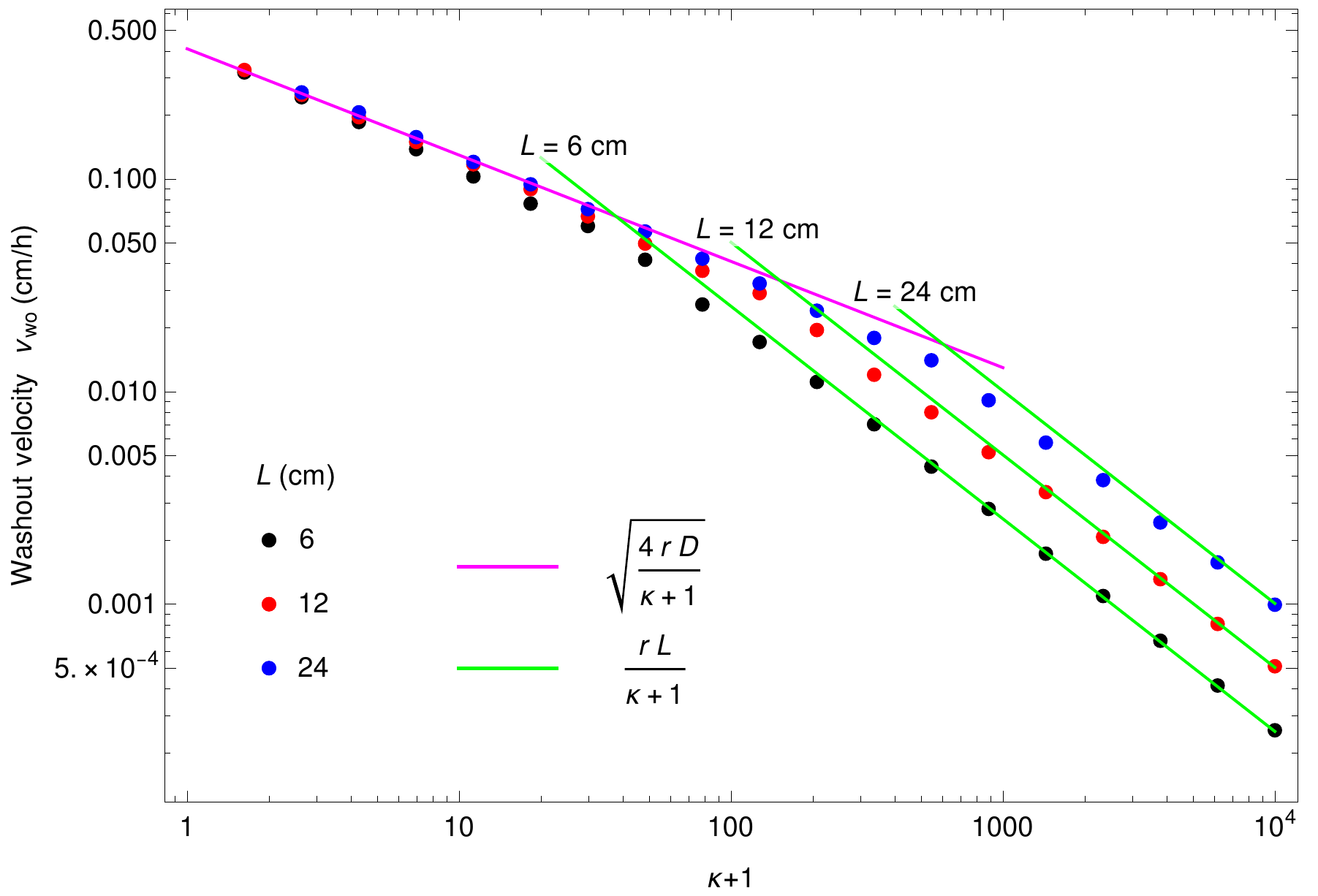}
\caption{
\textbf{Washout limits in the spatial system as a function of $\kappa+1$.} The two washout limits are obtained by fixing the diffusion coefficient at  \SI[per-mode=symbol]{0.1}{\centi\meter\squared\per\hour}, and varying the velocity for 20 different values of $\kappa+1$. The first velocity for which $F(L)>1.-10^{-10}$   is recorded as the washout velocity $v_{\text{wo}}$.
The process is repeated for three different values of length $L=6,12,24$ \SI{}{\centi\meter}.
Other parameters are  $v\:\fin=$ \SI[per-mode = symbol]{1}{\milli\Molar\centi\meter\per\hour}, $k=\kappa \fin$ \SI[per-mode = symbol]{}{\milli\Molar}, $r=$ \SI{0.42}{\per\hour}, $\alpha=$ \SI[per-mode = symbol]{6.13e8}{\bacteria\per\milli\Molar}. The green and magenta lines correspond to the two washout limits described by Eqs.~\ref{eq:WO1} and~\ref{eq:WO2}, respectively.}\label{fig:WOL}
\end{figure}

It is possible to find analytical estimates for the washout limits in the spatial system by comparing key length scales and time scales, as discussed in Ref.~\cite{Cremer2016}.  Let us first consider the case of large diffusion coefficients, when concentration profiles are flat regardless of the value of $v$. We compare the time $\tau_{\text{flow}}$ needed for a bacterium to travel through the system from the entrance to the exit to the minimal time $\tau_{\text{repl}}$ taken by a bacterium to replicate in this system (for $F=F_{in}$). If $\tau_{\text{flow}} < \tau_{\text{repl}}$, i.e.
\begin{equation}
		\frac{L}{v} < \frac{\kappa+1}{r},
\end{equation}
bacteria get washed out, so that an estimate of the washout velocity is
\begin{equation}\label{eq:WO1}
	v_{\text{wo}} = \frac{rL}{\kappa+1},
\end{equation}
which is in good agreement with the numerical results, as shown by Figure~\ref{fig:WOL}, and Figure~\ref{fig1}C in the main text.

Next, let us compare the diffusion and flow characteristic lengths at the time of replication, $\tau_{\text{repl}}$. Washout occurs if $L_{\text{diff}}<L_{\text{flow}}$, i.e. 
\begin{align}
	\sqrt{2 D \tau_{\text{repl}}}&<v\tau_{\text{repl}},\\
	\sqrt{2 \frac{D(\kappa+1)}{r}} &< \frac{v(\kappa+1)}{r},
\end{align}
which gives
\begin{equation}
	v_{\text{wo}} = \sqrt{\frac{2rD}{\kappa+1}}.
\end{equation}
This second washout limit is in agreement with Eq.~\ref{vstar}, as well as with the numerical results depicted in Figures~\ref{fig:WOL} and \ref{fig1}, up to a factor 2. Indeed, numerically, we find a good agreement with
\begin{equation}\label{eq:WO2}
    v_{\text{wo}} =v^*= \sqrt{\frac{4rD}{\kappa+1}}.
\end{equation}

The results in Figure~\ref{fig:WOL} are consistent with those of Figure~\ref{fig1}C where $(\kappa+1)\leq1.25$, and demonstrate the robustness of these results across a wide range of $\kappa$ values.

\clearpage

\section{Correspondence between the spatial system and the chemostat}\label{SM:chemostat}

\subsection{Main matching condition}\label{SM:main_match}

We wish to match the total number of divisions occurring in the spatial system and in the chemostat. At stationary state we thus aim to match the amount of food entering and exiting these different systems, as well as the amount of bacteria exiting them.

\begin{itemize}
\item Spatial system: amount of food entering: $dN_\text{F,in}/dt=\fin  v S$; amount of food exiting: $dN_\text{F,out}/dt=F(L) v S$; amount of bacteria exiting: $dN_\text{B,out}/dt=B(L) v S$;
\item Chemostat: amount of food entering: $dN_\text{F,in}/dt=\fin^\ch cV$; amount of food exiting: $dN_\text{F,out}/dt=F^\ch cV$; amount of bacteria exiting: $dN_\text{B,out}/dt=B^\ch cV$.
\end{itemize}
Here $V$ is the volume of the chemostat, $c$ is the dilution rate i.e. the outflow rate per unit volume of the chemostat. Concentrations in the chemostat are indicated by a superscript $\ch$. Meanwhile, $v$ denotes the velocity in the spatial system, $S$ the section and $L$ the length of the spatial system.

Hence, our matching condition reads:
\begin{equation}
\frac{\fin}{\fin^\ch}=\frac{F(L)}{F^\ch}=\frac{B(L)}{B^\ch}=\frac{cV}{vS}\,.
\label {match}
\end{equation}

\subsection{Constraints from each separate system}\label{SM:constr}

In the spatial system, we have $F(L)=\fin \phi(\kappa,\lambda,\sigma)$ and $B(L)=\alpha\left[\fin -F(L)\right]$, i.e. $B(L)=\alpha \fin \left[1-\phi(\kappa,\lambda,\sigma)\right]$, where $\phi$ is the dimensionless function introduced above.

In the chemostat, the following equations are satisfied:
\begin{align}
\frac{d F^\ch}{dt}&=-\frac{r}{\alpha}\frac{F^\ch B^\ch}{k+F^\ch}+c\fin^\ch-c F^\ch \,,\label{chemo1}\\
\frac{dB^\ch}{dt}&=r\frac{F^\ch B^\ch}{k+F^\ch}-c B^\ch \,. \label{chemo2}
\end{align}
At stationary state 
\begin{align}
0&=-\frac{r}{\alpha}\frac{F^\ch B^\ch}{k+F^\ch }+c\fin^\ch -cF^\ch \,,\\
0&=r\frac{F^\ch B^\ch }{k+F^\ch }-cB^\ch \,.
\end{align}
If $B^\ch \neq 0$, this yields $rF^\ch /(k+F^\ch )=c$, which means that the dilution rate $c$ of the chemostat sets the effective division rate of the bacteria, a fundamental chemostat property. And then (with $c\neq0$)
\begin{align}
F^\ch &=\frac{kc}{r-c}\,,\\
B^\ch &=\alpha\left(\fin^\ch -F^\ch \right)=\alpha \fin^\ch \left(1-\frac{k}{\fin^\ch }\frac{c}{r-c}\right)\,.
\end{align}

Hence, our matching condition Eq.~\ref{match} reads:
\begin{equation}
\frac{\fin}{\fin^\ch }=\frac{\fin\phi(\kappa,\lambda,\sigma)}{kc/(r-c)}=\frac{\alpha \fin \left[1-\phi(\kappa,\lambda,\sigma)\right]}{\alpha \fin^\ch \left[1-k/\fin^\ch \, c/(r-c)\right]}=\frac{cV}{vS}\,,
\end{equation}
which reduces to two equations relating the spatial system (left hand-side) to the chemostat (right hand-side):
\begin{align}
\phi(\kappa,\lambda,\sigma)&=\frac{k}{\fin^\ch }\frac{c}{r-c}\,,\label{matchb}\\
v S \fin &=cV\fin^\ch \label{matchc}\,.
\end{align}

We  assume that the parameters of the spatial system are given. Then we need to choose those of the chemostat in order to have a good matching. The parameters specific to the chemostat are $c,\fin^\ch ,V$. Note that $k$ and $r$ are assumed to be the same in both systems. In principle Eqs.~\ref{matchb} and~\ref{matchc} allow us to fix 2 out of these 3 free chemostat parameters. 

\subsection{Additional matching conditions}\label{SM:add_match}
We may want to impose additional matching conditions between the chemostat and the spatial system:
\begin{enumerate}
\item Same total volume: $V=SL$. This implies $\frac{cV}{vS}=\frac{cL}{v}$, and Eq.~\ref{match} would be modified accordingly.
\item Same volume exiting per unit time: $cV=vS$. This implies $\frac{cV}{vS}=1$, and Eq.~\ref{match} would be modified accordingly.
\item Same outflow rate relative to the total volume: $c=\frac{vS}{LS}=\frac{v}{L}$. This implies $\frac{cV}{vS}=\frac{V}{LS}$, and Eq.~\ref{match} would be modified accordingly.
\end{enumerate}
We note that if we impose two of these three conditions simultaneously, then the third one is also satisfied automatically, but Eqs.~\ref{matchb} and~\ref{matchc} and all three conditions above can be satisfied simultaneously only when $F(L)=vk/(rL-v)$.

\subsection{Properties of the matching chemostats}\label{SM:propert}
\begin{figure}[h!]
\includegraphics[width=\textwidth]{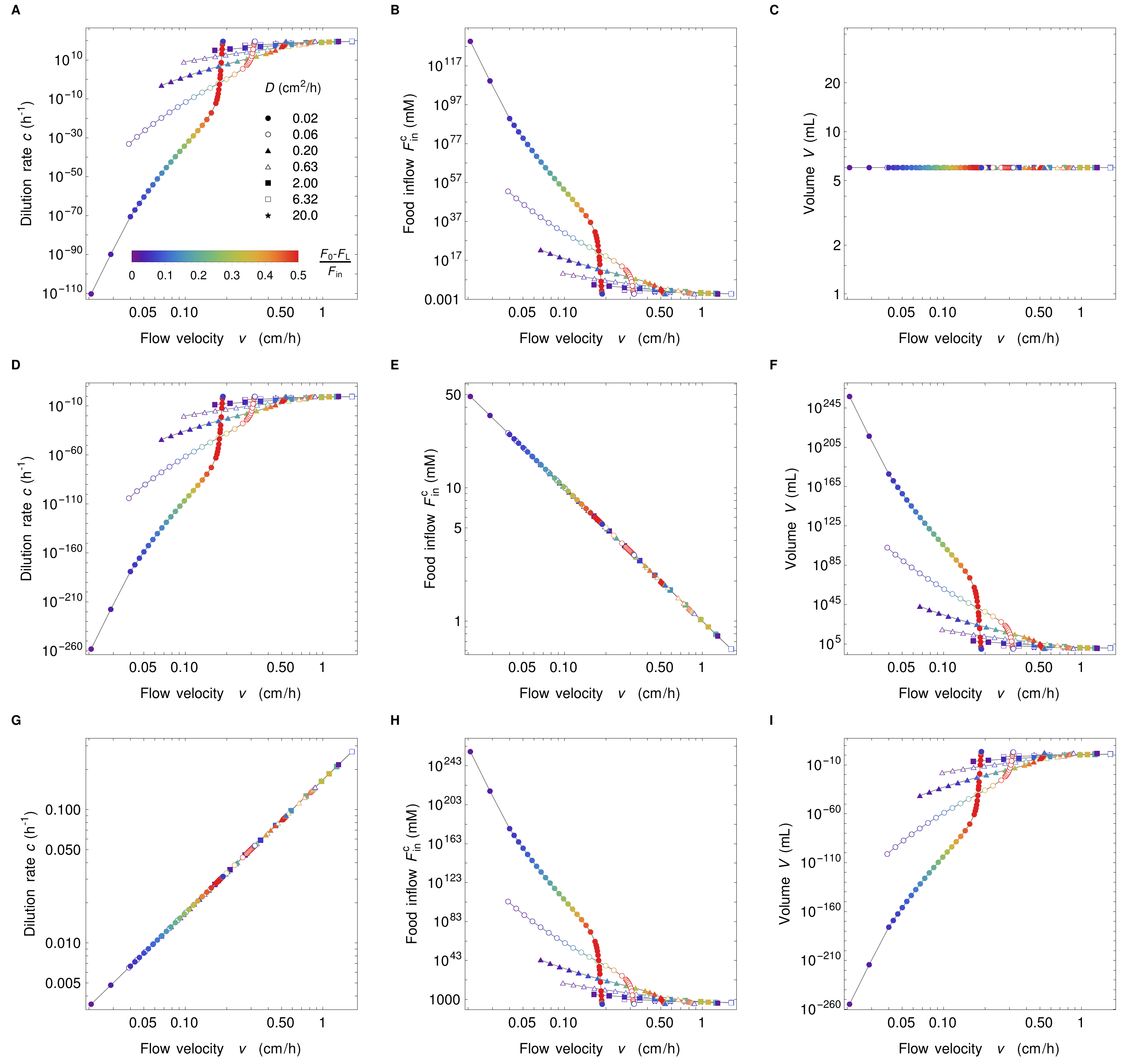}
\caption{
\textbf{Chemostat parameters.}  Parameters of the chemostat matching the spatial system in the conditions of Figure~\ref{fig3}. Each row in the figure  (top to bottom) represents matching condition 1 to 3, while each column in the figure represents a given parameter: dilution rate $c$ (left), food inflow $\fin       ^\text{c}$ (middle), and volume $V$ (right). Due to the very small values of food concentration exiting the gut, parameters of the chemostat system can have very large or very small values. This is particularly true for smaller diffusion constants and/or small velocities. }\label{fig:chem_param}
\end{figure}

In general, we can impose only 3 independent conditions, setting the values of $c,\finc,V$. Specifically, we have to take Eqs.~\ref{matchb} and~\ref{matchc}, plus one of the three conditions numbered 1, 2 and 3 above. These three possibilities are discussed in Table~\ref{tb:conditions} and illustrated in Figure \ref{fig:chem_param}.

\begin{small}
\begin{table}[htb!]
\makegapedcells
	\centering
\begin{tabular}{|p{2.8cm}|p{1.6cm}|p{4.1cm}|p{3.9cm}|p{2.6cm}|}
	\hline
	\multicolumn{2}{|p{4.8cm}|}{Matching condition} & $c$ & $\finc$ & $V$\\ \hline
	 1. Same total volume 
	 	& \( \displaystyle \frac{cV}{vS}=\frac{cL}{v}  \) 
	 	& \( \displaystyle \frac{F(L) v}{2Lk}\left(\sqrt{\frac{4Lrk}{F(L)v}+1}-1\right) \) 
		& \( \displaystyle \frac{\fin v}{2Lr}\left(1+\sqrt{\frac{4Lrk}{F(L)v}+1} \,\right)\) 
	 	& \( \displaystyle L\:S \) \\ \hline
	 2. Same volume exiting per unit time 
	 	& \( \displaystyle \frac{cV}{vS}=1  \) 
	 	& \( \displaystyle \frac{F(L)r}{k+F(L)}  \) 
		& \( \displaystyle \fin  \)  	 	
	 	& \( \displaystyle \frac{v\:S(F(L)+k)}{F(L)r} \)\\ \hline
	 3. Same outflow rate relative to the total volume 
		& \( \displaystyle \frac{cV}{vS} =\frac{V}{S L}  \)  
		& \( \displaystyle \frac{v}{L}  \)
		& \( \displaystyle \frac{\fin kv}{F(L)(Lr-v)} \)  
		& \( \displaystyle \frac{F(L)\:L\:S(Lr-v)}{kv} \) \\ \hline
\end{tabular}
\caption{Chemostat parameters $c, \finc, V$ as a function of the parameters of the spatial system for three different matching conditions. }\label{tb:conditions}
\end{table}
\end{small}

\clearpage

Finally, table~\ref{tb:TPR} gives the expression of various useful quantities for the spatial system and for the chemostat.

\begin{table}[htb!]
\makegapedcells
	\centering
\begin{tabular}{|p{.175\textwidth}|p{.45\textwidth}|p{.25\textwidth}|}
	\hline
	& Spatial system & Chemostat \\ \hline
	Reproduction rate  & 
		\( \displaystyle \rho(x) = r\frac{F(x)}{k + F(x)} \) & 
		\( \displaystyle \rho^\ch = r\frac{F^\ch}{k + F^\ch} \) \\ \hline
	\multirow{2}{=}{Reproductions per unit volume and unit time} &
		\(\displaystyle R(x) = B(x)\rho(x) \) 
			& \(\displaystyle R^\ch = B^\ch \rho^\ch  \) \\
		& \(\displaystyle \hphantom{R(x)} = r\alpha\frac{F(x)\left[\fin-F(x) \right]}{k + F(x)}\) 
			& \(\displaystyle \hphantom{R^\ch} = r\alpha  \frac{F^\ch\left(\finc-F^\ch\right)}{k +F^\ch} \)\\ \hline
	\multirow{2}{=}{Total reproduction rate } &  
		    \(\displaystyle N_\text{R} = r\alpha\:S\int_0^L\frac{F(x)\left[\fin-F(x)\right]}{k+F(x)}dx \) & 
			\(\displaystyle N_{\text{R}}^\ch   =  r\alpha V \frac{F^\ch\left(\finc-F^\ch\right)}{k +F^\ch}  \)\\
		 &  \(\displaystyle \hphantom{N_\text{R}} =  \alpha v  S\left[ \fin-F(L)  \right] \) &
		    \(\displaystyle  \hphantom{N_\text{R}^\ch}=  \alpha c  V\left(\finc-F^\ch  \right) \) \\ \hline
	Total population  & 
		\(\displaystyle  N_{\text{T}}  =  \alpha S \int_0^L\left[\fin-F(x)\right]dx \) & 
		\(\displaystyle N_{\text{T}}^\ch  = \alpha  V \left(\finc-F^*\right)  \) \\ \hline
	Active population  & 
		\(\displaystyle N_\text{A}=\alpha  S \int_0^{x^*}\left[\fin-F(x)\right]dx , \quad x^*: F(x^*)=k  \) & 
		\(\displaystyle   N_{\text{A}}^\ch=N_{\text{T}}^\ch  \) \\ \hline
	Fixation probability &
		\(\displaystyle \mathcal{F} = \frac{\int_0^L R(\xM) \frac{M(\xM)}{B(\xM)} d\xM }{\int_0^L R(\xM)d\xM} \) &
		\(\displaystyle \mathcal{F}^\ch = \frac{N_\text{M}}{N_\text{T}^\ch} \)	\\ \hline
	Washout limit  & \(\displaystyle v_\text{wo} = \begin{cases} \frac{rL}{k/\fin+1}\\ \sqrt{\frac{4rD}{k/\fin+1}}\end{cases} \) & \(\displaystyle c=\frac{r}{k/\finc+1} \) \\ \hline
	\end{tabular}
	\caption{Comparison of the main derived quantities in the spatial system and the chemostat. 
	}\label{tb:TPR}
\end{table}

\clearpage

\section{Early dynamics of mutant bacteria concentration}\label{SM:mut_dynam}

\begin{figure}[h!]
\includegraphics[width=.93\textwidth]{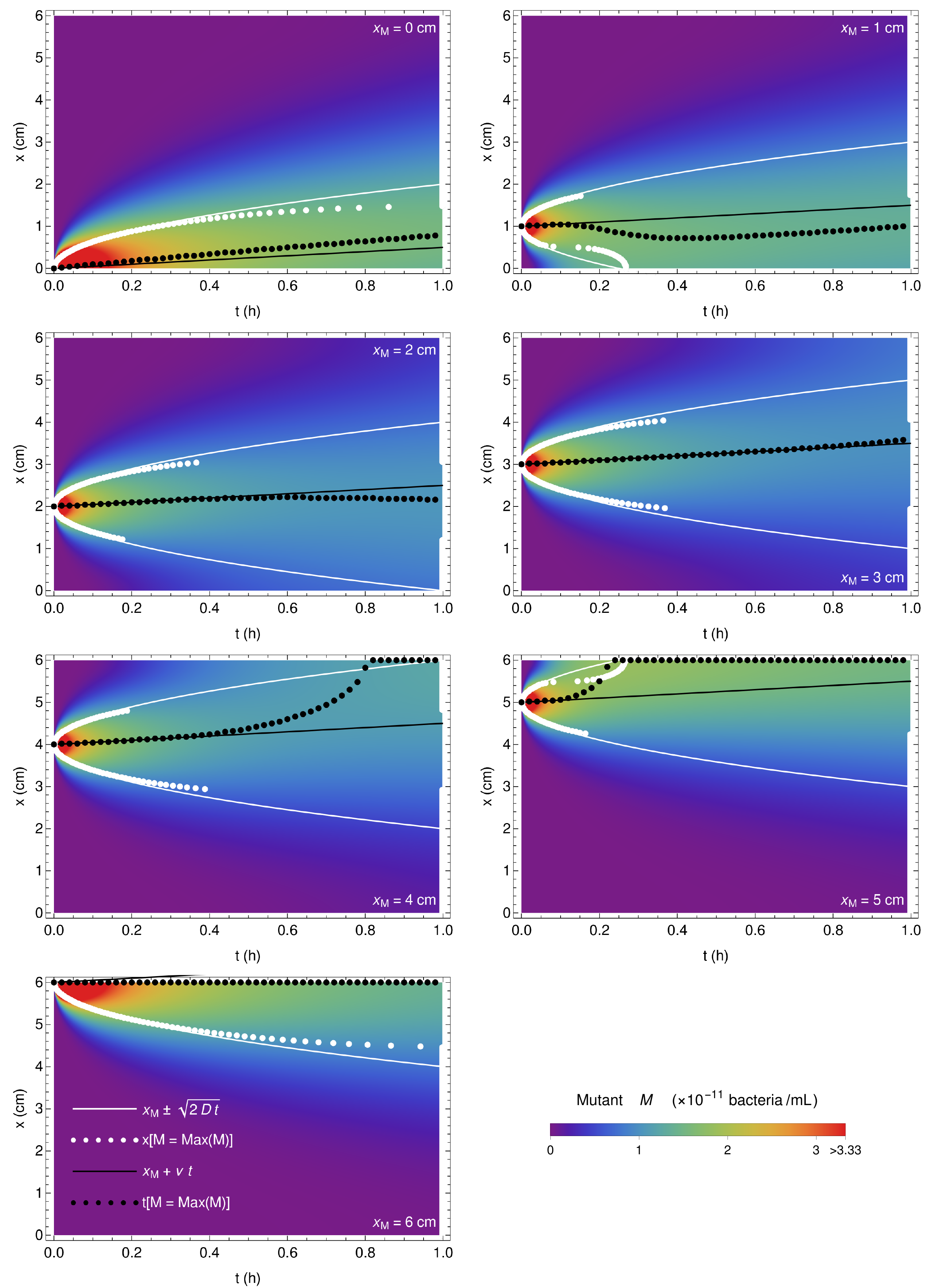}
\caption{
\textbf{Early dynamics of mutant concentration.} Spatio-temporal evolution of mutant concentration for $\xM = 0,\: 1,\: 2,\: 3,\: 4,\: 5,\:$\SI{6}{\centi\meter} in the first hour after mutant introduction. Black points correspond to the maximum mutant concentration at time $t$, white points correspond to the maximum mutant concentration at position $x$, black line is $\xM+vt$, and white curve is $\xM\pm\sqrt{2Dt}$.
The parameter values are $D=$ \SI[per-mode = symbol]{0.4}{\centi\meter\squared\per\hour}, $v=$ \SI[per-mode = symbol]{0.5}{\centi\meter\per\hour}, $k=$ \SI[per-mode = symbol]{0.1}{\milli\Molar}, $r=$ \SI{0.42}{\per\hour}, $v\:\fin =$ \SI[per-mode = symbol]{1}{\milli\Molar\centi\meter\per\hour}, $\alpha=$ \SI[per-mode = symbol]{6.13e8}{\bacteria\per\milli\Molar} and $M_0=3.33\times10^{-9}$\SI[per-mode = symbol]{}{bacteria\per\milli\liter}. }\label{fig:MutDynam}
\end{figure}

\clearpage

\section{Stationary state of mutant bacteria concentration versus the initial position $\xM$ of mutants}\label{SM:mut_stat_state}

\begin{figure}[h!]
\includegraphics[width=\textwidth]{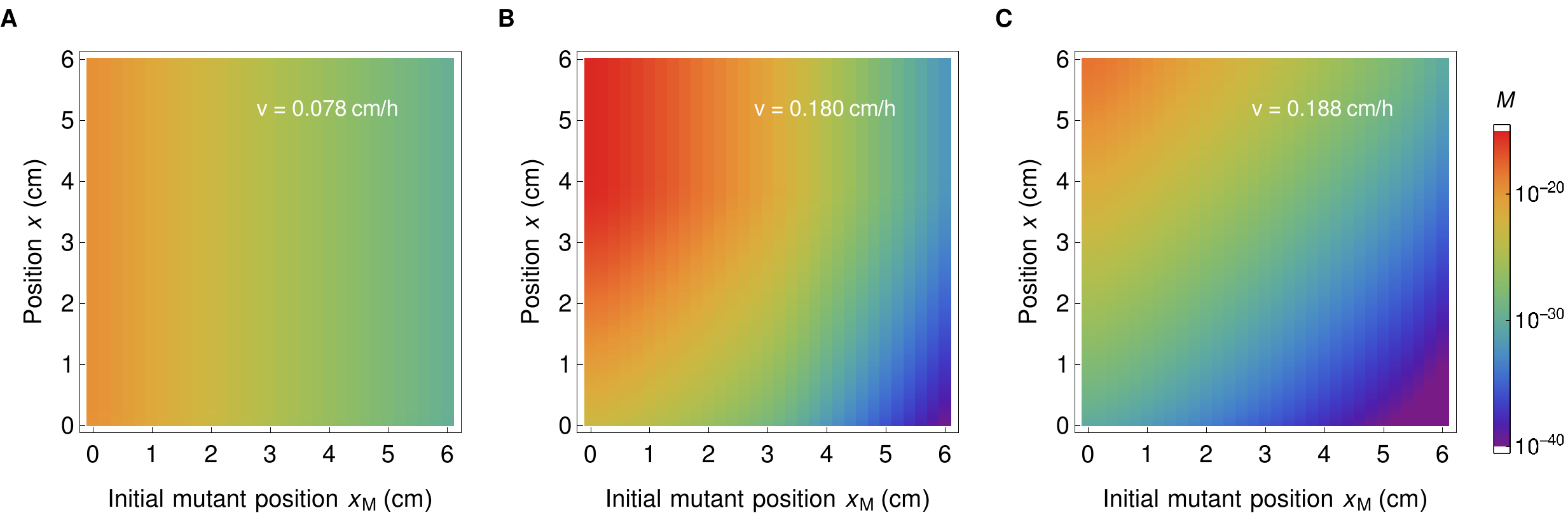}
\caption{
\textbf{Stationary distribution of mutant bacteria in the gut.} Concentration of mutant bacteria as a function of position $x$, and initial mutant position $\xM$, for $D=$\SI[per-mode = symbol]{0.02}{\centi\meter\squared\per\hour}, and for three different values of flow velocity $v$, corresponding to the flat concentration profile in the well mixed regime (\textbf{A}), spatial concentration profile (\textbf{B}), and close to the washout limit (\textbf{C}). In all three cases, we observe that the final concentration of the mutant bacteria is smaller if the initial position $\xM$ is further along the gut.  This figure corresponds to three points of the top curve in Fig.~\ref{fig3}.}\label{fig:M(x,xM)}
\end{figure}

\begin{figure}[h!]
\includegraphics[width=\textwidth]{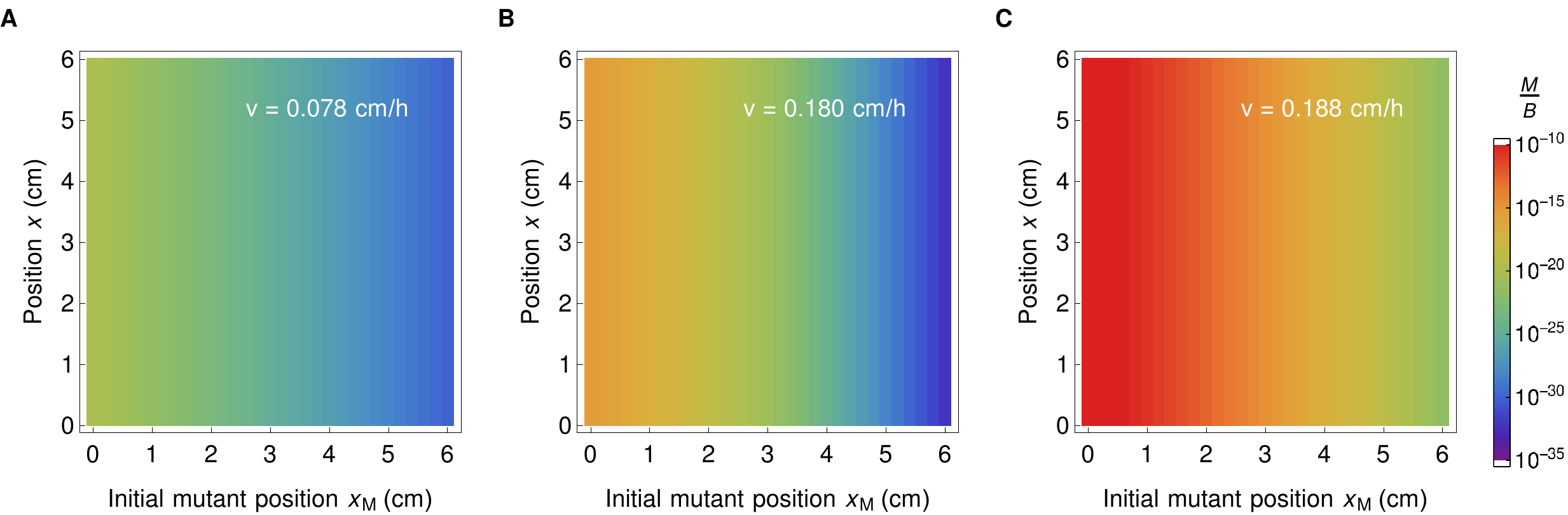}
\caption{
\textbf{Stationary ratio of mutant to wild type bacteria concentrations.} The ratio $M/B$ is shown in all three cases corresponding to Fig.~\ref{fig:M(x,xM)}. Consistently with our analytical predictions discussed in the main text, the $M/B$ ratio is constant along the gut. In addition, we observe that it monotonically decreases as a function of initial mutant position $\xM$, and the decrease is the most pronounced in the case of the spatial profile (panel \textbf{B}) where the ratio at the beginning and at the end of the gut are several orders of magnitude apart. }\label{fig:MB(x,xM)}
\end{figure}

\clearpage

\section{Location where most mutants that fix originate}\label{SM:maxR}

\begin{figure}[h!]
\includegraphics[width=\textwidth]{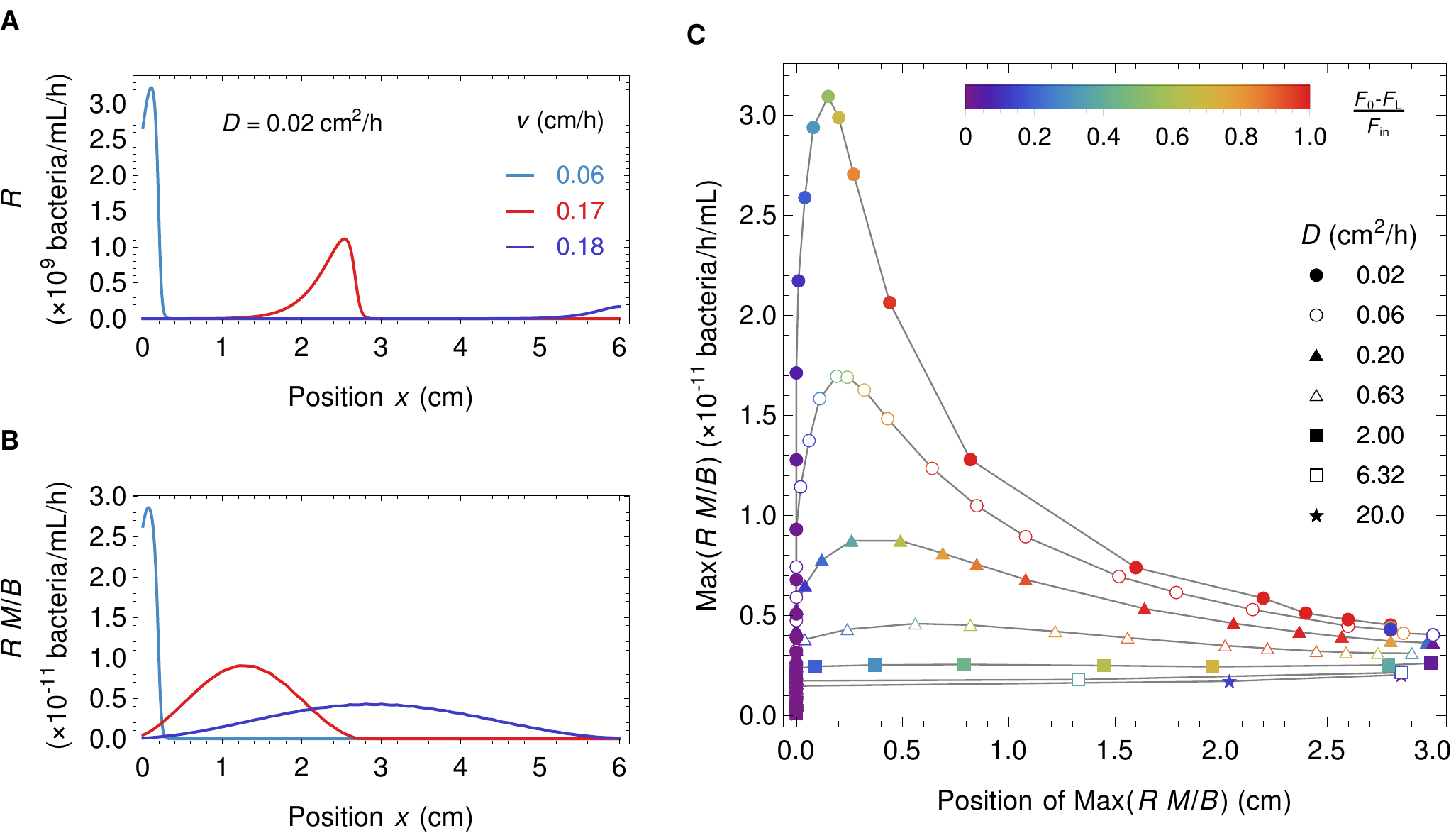}
\caption{
\textbf{Maximum of $R$ and $R\: M/B$.} 
\textbf{A:} Number $R$ of reproduction events per unit volume and unit time as a function of position along the gut for three different types of concentration profiles, almost well-mixed (light blue), spatial (red), and close to the washout limit (purple).
\textbf{B:} Maximum of the product $R\:M/B$ of the reproduction events per unit volume and unit time and of the ratio of mutant and wild type bacteria as a function of space for the same three concentration profiles as in A (same colors).
\textbf{C:} Maximum value of $R\:M/B$ as a function of its position for the data set in Figure~\ref{fig3} of the main text. 
}\label{fig:maxR}
\end{figure}

While the position of the maximum of the ratio $M/B$ of mutant to wild type bacteria concentrations is always at the entrance of the gut (see main text and Figure~\ref{fig2}), the position of the maximum of the number $R$ of reproduction events per unit volume and unit time depends on parameter values. We find that for flat concentration profiles, it is located either close to the entrance of the gut (for small velocities yielding an almost well-mixed system) or at the exit of the gut (close to the washout limit). Conversely, for spatial concentration profiles, its location is intermediate (see Figure~\ref{fig:maxR}A). Because of this, in the regime with strong spatial dependence, we find that the position of the maximum of the product $R\:M/B$ of these two quantities ranges between 0 and $L/2$ (see Figure~\ref{fig:maxR}B and C). The position of the latter maximum corresponds to the location where most mutants that fix tend to originate.

\section{Calculation of the active population size}\label{SM:active_pop}

The active population corresponds to the bacteria in the zone where reproduction rate is significant. Concretely, it is defined as
\begin{equation}
    N_{\text{A}} = S \int_0^{x^*} B(x) dx\,,
\end{equation}
where $x^*$ corresponds to the point in the gut segment where the food concentration equals the Monod constant $k$, i.e. $F(x^*)=k$, implying that the reproduction rate $\rho(x^*)$ is equal to half of the maximal possible reproduction rate, which is obtained if $F(x)\gg k$. Thus,
\begin{equation}
    B(x^*) = \alpha [\fin-F(x^*)] = \alpha [\fin-k] = \alpha\fin(1-\kappa)\,.
\end{equation}
In the active population thus defined, the reproduction rate of bacteria is at least equal to half of its maximal possible value.

\clearpage

\section{Impact of the dimensionless parameters}\label{SM:param}

\subsection{Holding the dimensionless parameters fixed}\label{SM:param_fixed}

\begin{figure}[h!]
\includegraphics[width=\textwidth]{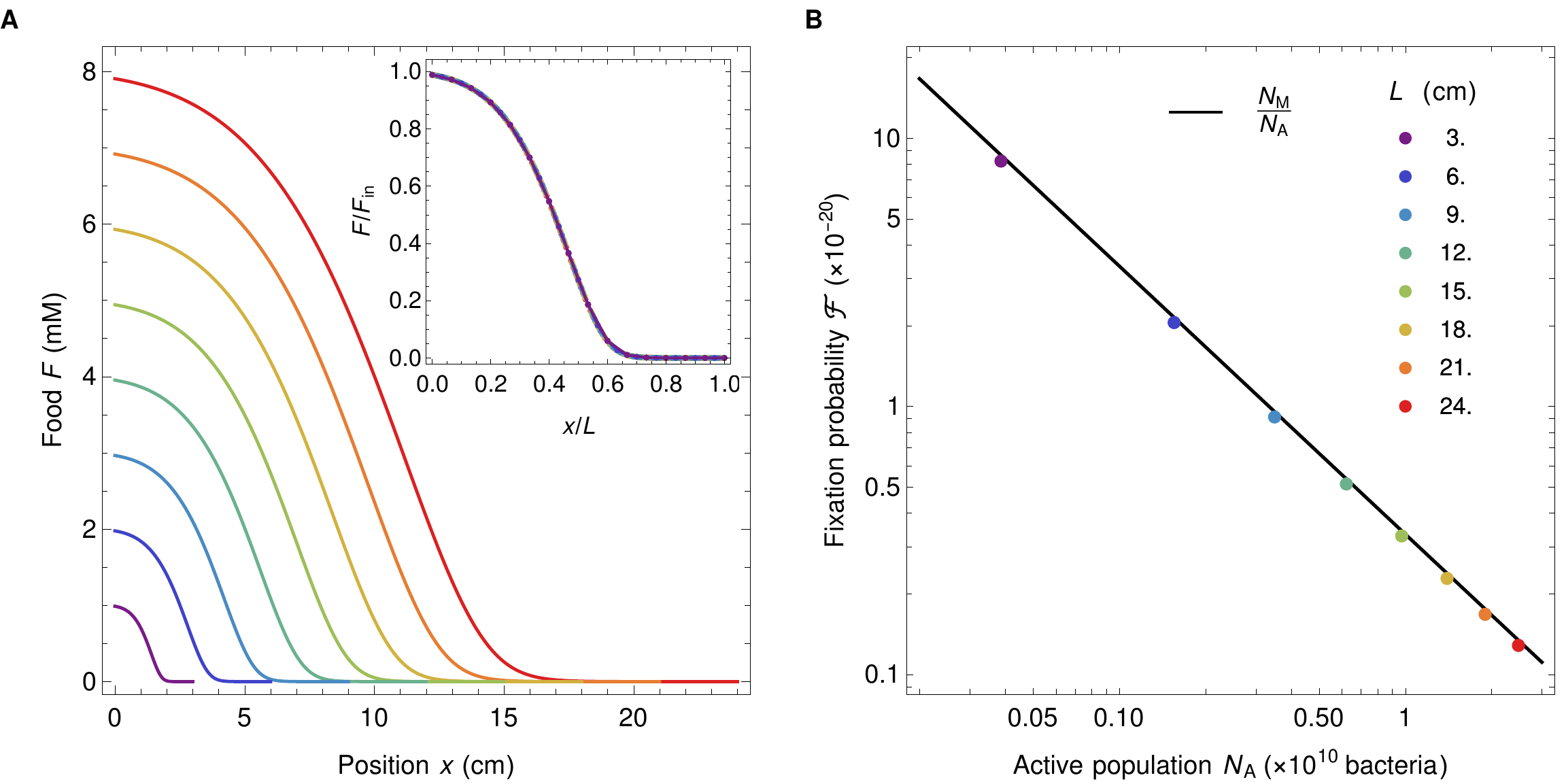}
\caption{
\textbf{} \textbf{A} Food concentration profiles for different system sizes and keeping nondimensional parameters, $\lambda, \kappa, \sigma$, fixed. The profiles can be rescaled  by dividing food concentration by $\fin$ and space by $L$ (see inset of A) showing all the profiles are identical up to the scaling factors. \textbf{B} Fixation probability for the profiles in panel \textbf{A}.  The parameters are $D=$ \SI[per-mode=symbol]{0.2}{\centi\meter\squared\per\hour}, $v=3./L$ \SI[per-mode=symbol]{}{\centi\meter\per\hour}, $\fin=1./v$ \SI{}{\milli\Molar}, $k=0.05/v$ \SI{}{\milli\Molar}, $r=1.68\cdot v^2$ \SI{}{\per\hour}, $\alpha=$ \SI[per-mode = symbol]{6.13e8}{\bacteria \per\milli\Molar}, $N_{\text{M}}=$ \SI{3.33e-11}{\bacteria} and the $L$ values are listed in the panel \textbf{B}.}\label{fig:NPF}
\end{figure}

To illustrate the relevance of the dimensionless parameters introduced in section~\ref{SM:dimless} to describe the stationary profiles, we vary system parameters so that we hold the dimensionless ones fixed. The reference for fixing them is Figure~\ref{fig2}:
\begin{align}
	\kappa	& = \frac{k}{\fin} = \frac{0.1}{2.0} = 0.050,\\
	\lambda	& = \frac{r D}{v^2} = \frac{0.42 \times 0.2}{0.5^2} = 0.34, \\
	\sigma	& = \frac{L v}{D} = \frac{6 \times 0.5}{0.2} = 15.
\end{align}
We first vary the gut length $L$ in order to ease the discretization of the space. Once the $L$ is chosen, the dimensionless parameters are fixed by adjusting $v$, and then $\fin$, to keep the product $v\:\fin$ constant. Other parameters are kept fixed. Figure~\ref{fig:NPF}A shows the corresponding concentration profiles. They all have the same shape, as evidenced by rescaling the food concentration by $\fin$ and the spatial coordinate by $L$ (see inset of the Figure~\ref{fig:NPF}A). Figure~\ref{fig:NPF}B shows that the fixation probability for these profiles scales with active population, $\mathcal{F}=N_{\text{M}}/N_{\text{A}}$, consistently with our expectations, since the concentration profiles are strongly spatially dependent. 

\newpage

\subsection{Varying each dimensionless parameter}\label{SM:param_vary}

\begin{figure}[h!]
\includegraphics[width=\textwidth]{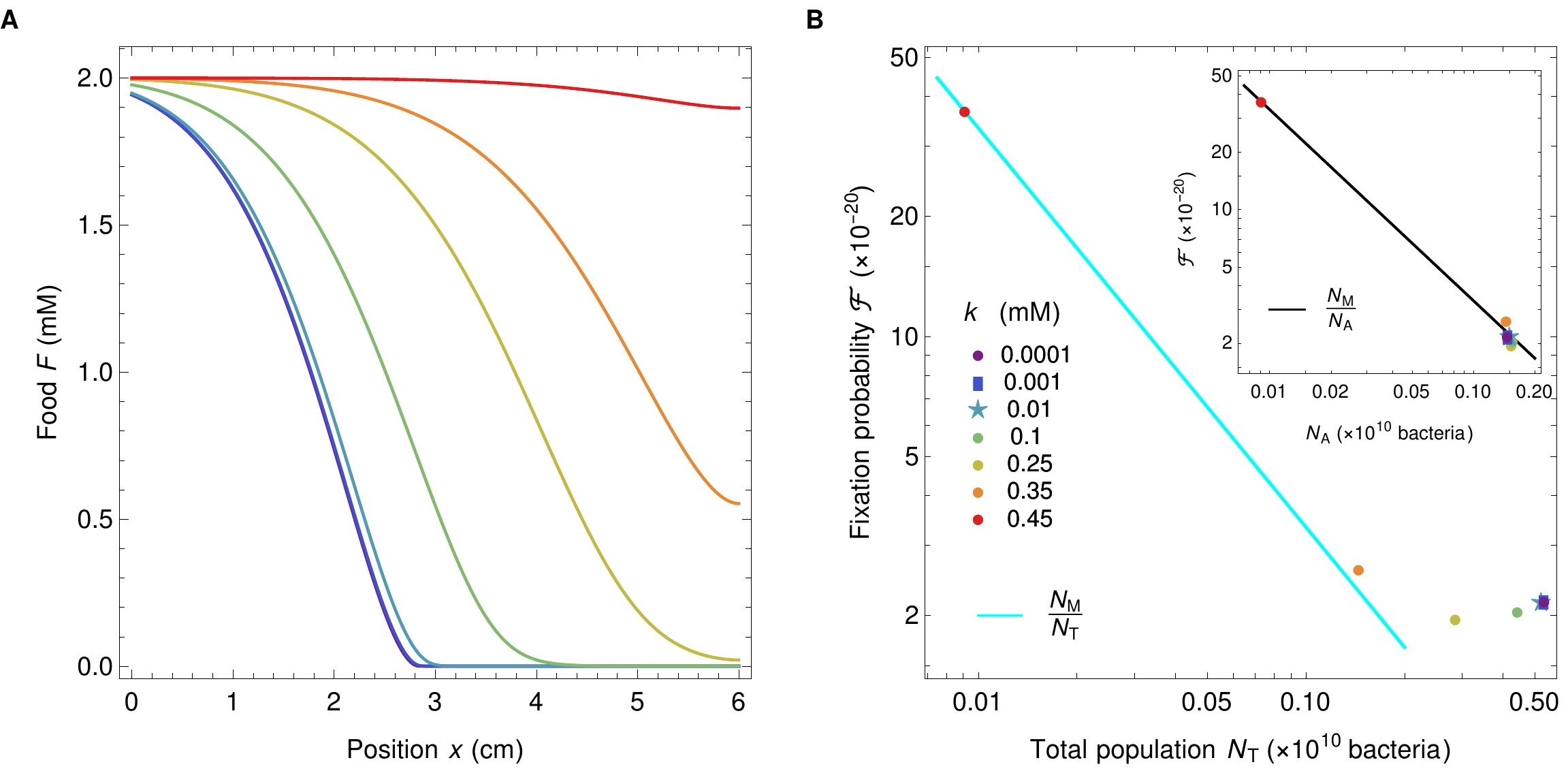}
\caption{
\textbf{Varying $\kappa$.} \textbf{A} Food profiles for different values of $\kappa$ and fixed $\lambda=0.336$ and $\sigma=15$. \textbf{B} Fixation probability as a function of total population $N_\text{T}$ and active population $N_\text{A}$ (in the inset). Different colors correspond to different values of $\kappa$ ($k$) and different symbols are used for overlapping data points. 
Parameters are $v=$~\SI[per-mode=symbol]{0.5}{\centi\meter\per\hour}, $D=$~\SI[per-mode=symbol]{0.2}{\centi\meter\squared\per\hour}, $L=$~\SI{6.}{\centi\meter}, $r=$~\SI{0.42}{\per\hour}, $\alpha=$~\SI[per-mode = symbol]{6.13e8}{\bacteria\per\milli\Molar}, and $N_{\text{M}}=$ \SI{3.33e-11}{\bacteria}. $k$ values are listed in the panel \textbf{B}.}\label{fig:NDPK}
\end{figure}

\begin{figure}[h!]
\includegraphics[width=\textwidth]{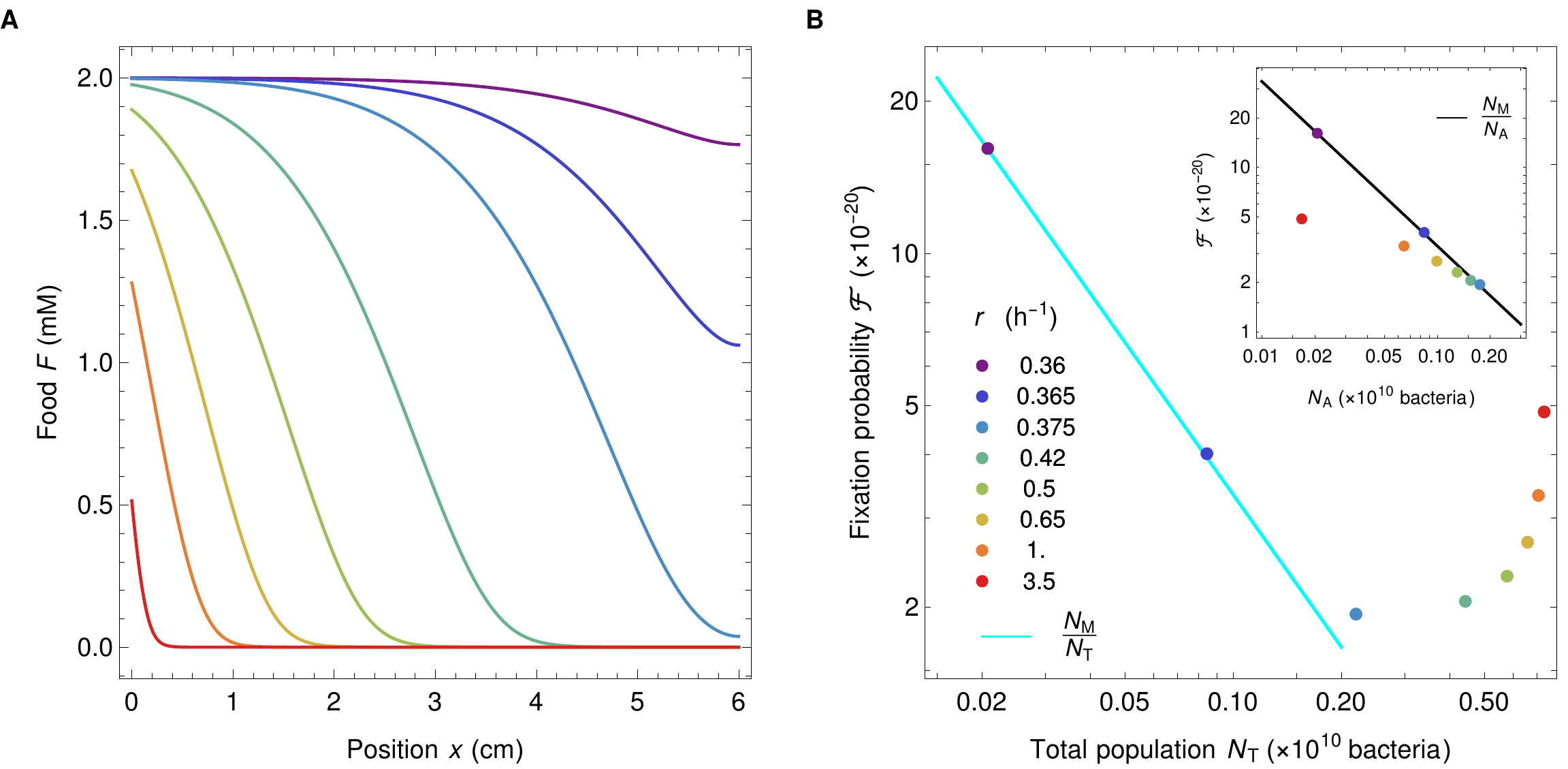}
\caption{
\textbf{Varying $\lambda$.} \textbf{A} Food profiles for different values of $\lambda$ and fixed $\kappa=0.05$ and $\sigma=15$. \textbf{B} Fixation probability as a function of total population $N_\text{T}$ and active population $N_\text{A}$ (in the inset). Different colors correspond to different vales of $\lambda$ ($r$).
Parameters are $v=$~\SI[per-mode=symbol]{0.5}{\centi\meter\per\hour}, $D=$~\SI[per-mode=symbol]{0.2}{\centi\meter\squared\per\hour}, $k=$~\SI{0.1}{\milli\Molar}, $k=$~\SI{0.1}{\milli\Molar}, $\alpha=$~\SI[per-mode = symbol]{6.13e8}{\bacteria \per\milli\Molar}, and $N_{\text{M}}=$ \SI{3.33e-11}{\bacteria}. $r$ values are listed in the panel \textbf{B}.}\label{fig:NDPR}
\end{figure}

\begin{figure}[h!]
\includegraphics[width=\textwidth]{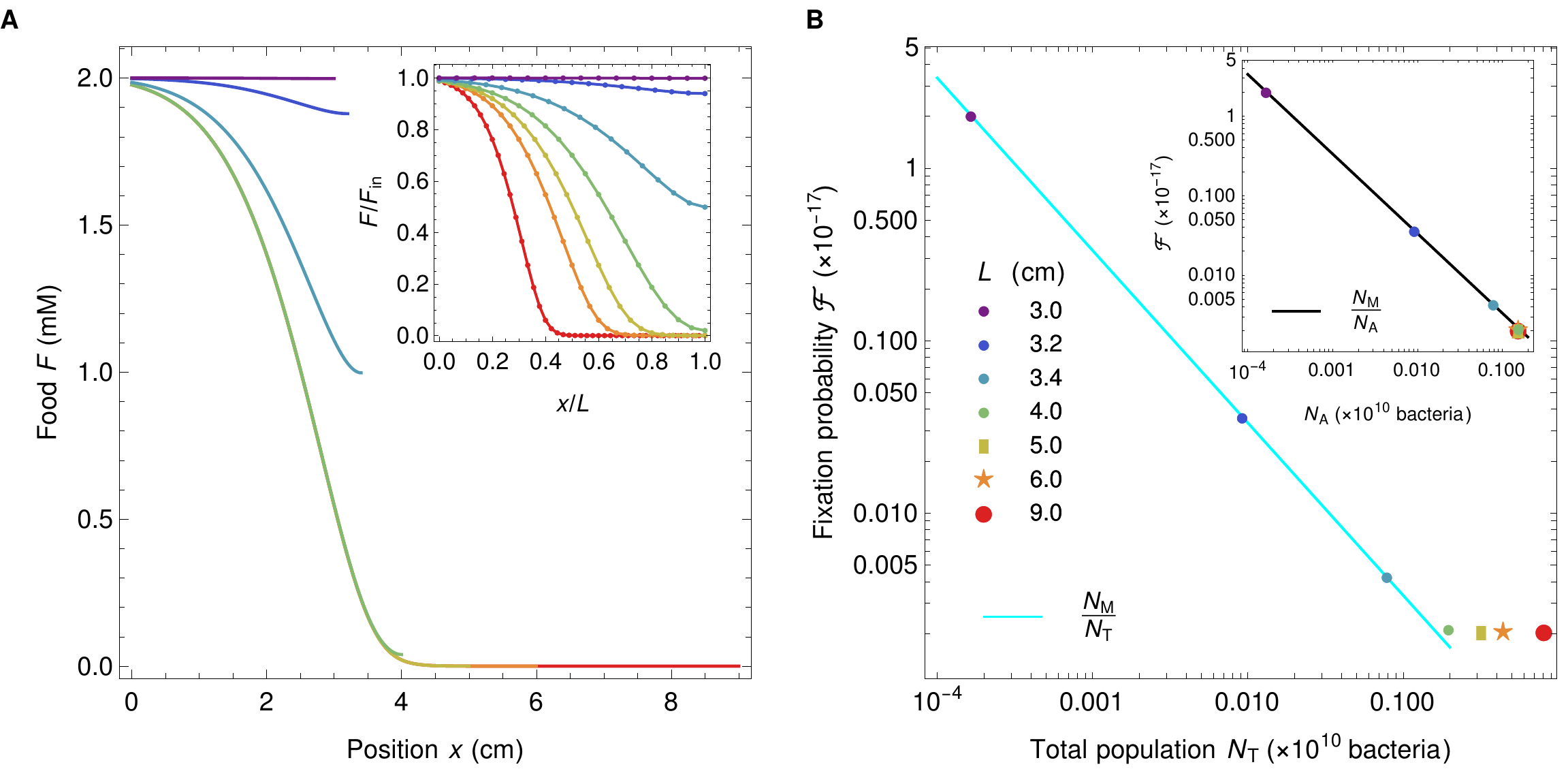}
\caption{
\textbf{Varying $\sigma$.} \textbf{A} Food profiles for different values of $\sigma$ and fixed $\lambda=0.336$ and $\kappa=0.05$. The inset shows rescaled food profiles, for easier comparison of the spatial dependence. \textbf{B} Fixation probability as a function of total population $N_\text{T}$ and active population $N_\text{A}$ (in the inset). Different colors correspond to different values of $\sigma$ ($L$) and different symbols are used for overlapping data points. 
Parameters are $v=$~\SI[per-mode=symbol]{0.5}{\centi\meter\per\hour}, $D=$~\SI[per-mode=symbol]{0.2}{\centi\meter\squared\per\hour}, $L=$~\SI{6.}{\centi\meter}, $r=$~\SI{0.42}{\per\hour}, $\alpha=$~\SI[per-mode = symbol]{6.13e8}{\bacteria\per\milli\Molar}, and $N_{\text{M}}=$ \SI{3.33e-11}{\bacteria}. $k$ values are listed in the panel \textbf{B}.}\label{fig:NDPL}
\end{figure}

Finally, we systematically investigate the impact of each dimensionless parameter by varying one of them while holding the other two fixed. We keep parameters $v$, $D$ and $\fin$ fixed throughout, and vary $k$, $r$, and $L$ one at a time in order to change $\kappa$, $\lambda$, and $\sigma$, respectively. Results in Figures~\ref{fig:NDPR}, \ref{fig:NDPK} and \ref{fig:NDPL} show that the fixation probability is well described by $\mathcal{F}=N_{\text{M}}/N_{\text{A}}$ except in cases where the food concentration profile is less spatial because food is substantially depleted even at the entrance of the gut, which occurs for the four largest values of $r$ in Figure~\ref{fig:NDPR}.

\subsection{Range of the dimensionless parameters and relevance for the human colon}

In this section, we discuss the range of values of the dimensionless parameters $\kappa$, $\lambda$ and $\sigma$, which are the only parameters of the model that may change the behavior of the system and affect our conclusions (recall that $\alpha$ is just a scaling factor that does not affect spatial profiles and conclusions, see section \ref{SM:OD_conversion}). We first summarize the range of values that these dimensionless parameters take in our study and in Refs.~\cite{Cremer2016,Cremer2017}. Next, we discuss the realistic range of values that they can take in the human colon, and show that the range studied here is relevant for the human colon. 

The parameter values employed in this paper correspond to the mini gut described in~\cite{Cremer2016}, where it has been proven that the mathematical model describes well the experimental setup. In~\cite{Cremer2017}, the same model was used to describe microbiota growth and composition in the human colon, and parameter values were thus altered to match the properties of the human gut, which is several times bigger than the mini gut. However, this change of parameter values did not significantly modify the values of the dimensionless parameters. As illustrated in Figure~\ref{fig:NPF}, holding the dimensionless parameters fixed fully preserves the properties  of the system, including the  spatial profiles of concentrations, and our conclusions on the active population remain true.
Table~\ref{tb:param_compare} summarizes the range of dimensionless parameter values considered in Figure~\ref{fig3} of this paper, and compares them to those of Refs.~\cite{Cremer2016,Cremer2017}. 

\begin{table}[htb!]
\makegapedcells
	\centering
\begin{tabular}{|p{2.2cm}|p{2.5cm}|p{3cm}|p{2.5cm}|p{2.5cm}|}
	\hline
	Dimensionless parameter	& Figure~\ref{fig3} & \mbox{Figure~\ref{fig3}};  $\frac{F(0)-F(L)}{\fin} > 0.9$ & Ref.~\cite{Cremer2016} & Ref.~\cite{Cremer2017} \\ \hline
		 \( \displaystyle \kappa  = \frac{k}{\fin}  \)	&$10^{-4}-0.21$ &$0.014-0.083$ &$0.05$ &$2.5\times10^{-4}$ \\ \hline
	 \( \displaystyle \lambda = \frac{rD}{v^2}  \)	&$0.19-8.4\times10^6$ & $0.24-0.42$ & $0.036-9.3$ & $0.29$ \\ \hline
	 \( \displaystyle \sigma  = \frac{Lv}{D}  \)	&$3\times10^{-4}-57$ &$7.59-56$ &$1.5-25$ & $9.6$ \\ \hline
\end{tabular}
\caption{Dimensionless parameter values used in Figure~\ref{fig3} of this paper and in Refs.~\cite{Cremer2016} and ~\cite{Cremer2017}.}\label{tb:param_compare}
\end{table}

Let us now discuss the realistic range of values of the dimensionless parameters $\kappa$, $\lambda$ and $\sigma$ in the human colon. Let us start by considering $\kappa=k/\fin$, and for this, let us first estimate $\fin$. Bacteria in the large intestine consume a mix of different nutrients that have not (or not completely) been absorbed in the small intestine. A large component of them are fibers, and a human typically ingests 25 to 100~g a day of fibers \cite{rose2015characterization}. The colon input also includes a smaller or similar quantity of unabsorbed sugars and starch \cite{mcneil1984contribution}. Given that the inflow of digesta in the colon is about 2 liters per day \cite{debongnie1978capacity}, the order of magnitude of the incoming food concentration $\fin$ in the colon is in the range of 15 to 100 g/L. The Monod constant $k$ depends on many factors, including the substrate and the bacterial strain. Even for glucose, it can range from 0.03 to 5 mg/L depending on how well adapted the bacteria are to growing on glucose~\cite{kovarova1998growth}, and it is typically higher for other substrates (20 to 300~mg/L for acetate~\cite{ramirez2009modeling}; 5 to 900 mg/L for different substrates~\cite{lawrence1969kinetics}). Realistic values for $\kappa=k/\fin$ are thus in a wide range, but in all cases, they are much smaller than 1, at least as small as 0.1. In the next section where dimensionless parameter values are systematically varied, Figure \ref{fig:NDPK} demonstrates that such values of $\kappa$ give very similar outcomes, and all results collapse in the limit $\kappa \to 0$.

The two other dimensionless parameters, $\lambda$ and $\sigma$, both involve the effective diffusion coefficient $D$, which models mixing and is thus hard to measure directly. However, it is empirically well demonstrated that there is a strong gradient of bacterial concentration along the colon \cite{GORBACH1967,Imai2018}, and that most nutrients that could be used by bacteria are consumed by the end of the gut \cite{mcneil1984contribution}. This requires that $\lambda= rD/v^2$ be larger than $1/4$ (washout limit, see section \ref{SM:washout}), and less than a few units (since above, diffusion is strong enough for the system to be almost well mixed). This can be seen on Figure 1C. Accordingly, Table \ref{tb:param_compare} demonstrates that despite the wide range of parameters used in our study, the range of $\lambda$ is very narrow in the regime that yields strongly spatial profiles (specifically, $\lambda$ is between 0.24 and 0.42 when $[F(0)-F(L)]/\fin>0.9$). This is in line with the estimate from \cite{Cremer2017}. 

Let us finally turn to $\sigma$. We notice that $\sigma =Lv/D= L r/(v \lambda)$. Avoiding washout requires $L/v>(\kappa +1)/r\gtrsim 1/r$ (see section \ref{SM:washout}), where the last inequality is rather tight because $\kappa \ll 1$. Since in addition 1/$\lambda$ is of the order of 2 to 4 in the very spatial regime, $\sigma = L r/(v \lambda)$ then has to be greater than a few units. Another way to estimate $\sigma$ is to employ $L r/v = \tau_\text{dig}/\tau_\text{rep}$ where $\tau_\text{rep}=1/r$ is the minimal bacterial replication time (within the colon), and $\tau_\text{dig} =L/v$ is the typical time spent in the system. While the total transit time ranges from one to several days \cite{cummings1976measurement}, the time spent by the digesta in the ascending colon, which is the upstream part of the colon where the strong gradients of food and bacterial concentrations are located~\cite{Cremer2017}, is substantially shorter, of the order of 4 hours \cite{proano1991unprepared}, and thus $\tau_\text{dig}\approx4$~h. Let us now estimate $\tau_\text{rep}$. Feces weigh about 130 g/day and contain 25--50\% of bacteria in mass \cite{rose2015characterization}. The typical mass of a bacteria being 1 pg, this means that about 3--6$\times 10^{13}$ bacteria per day are lost in feces, to be compared with about $4\times10^{13}$ bacteria in the colon \cite{sender2016}, leading to about one renewal per day, which means that $\tau_\text{rep}\leq 24$ h. However, as it is likely that bacteria actively replicate only in the upper part of the colon, while $r$ represents the maximal reproduction rate in the gut, the actual value of $\tau_\text{rep}$ is expected to be substantially smaller than this upper bond. A lower bound for $\tau_\text{rep}$ is given by the minimal doubling time of fast replicating bacteria such as \textit{Escherichia coli} in good conditions, which can be as low as 20 minutes \cite{gibson2018distribution}.  
To summarize, $\sigma = \tau_\text{dig}/(\lambda\tau_\text{rep})$, with $1/\lambda$ of the order of 2--4 and $\tau_\text{dig}/\tau_\text{rep}\gtrsim 1$, and $\tau_\text{dig}/\tau_\text{rep}\approx 10$ when considering the ascending colon and the maximal replication rate. This matches well the range of values of $\sigma$ considered in the present work (see Table~\ref{tb:param_compare}).

\section{Relevance of neutral mutations }
\label{neutrality}

In this study, we focused on neutral mutations, as a first step towards understanding evolution in the gut. What fraction of spontaneous mutations is expected to be effectively neutral in gut bacteria? Despite the importance of beneficial mutations for adaptation, they are expected to be a small fraction of spontaneous mutations, at least in reasonably well-adapted organisms. Therefore, let us introduce the fitness cost $\delta$ of a mutation, defined by $f_M/f_W=1-\delta$ where $f_M$ is mutant fitness and $f_W$ wild-type fitness (beneficial mutations then have a negative $\delta$). Mutations are effectively neutral if their fitness cost satisfies $N|\delta|\ll 1$ where $N$ is the effective population size~\cite{Ewens79}, which is the active population size here. 

Let us estimate active population size in the human colon, where the total bacterial population is of the order of $4 \times 10^{13}$ bacteria \cite{sender2016}. The total rate of bacterial divisions in the gut is given by the ratio of the total population size to the renewal time. This rate can also be expressed as the active population size divided by the minimal replication time $\tau_\text{rep}$, because bacteria in the active population reproduce at rates close to the maximal one. Thus, the ratio of active to total population size is equal to the ratio of minimal replication time to renewal time, which, as discussed in previous section, can be as small as $20/(24\times60)=1.4\times 10^{-2}$, and is likely to be of order 0.1 (also consistent with \cite{ghosh2021emergent}). In addition, fixation of a mutant occurs within a strain. Different strains may compete for resources, but here, for simplicity, we do not consider this aspect, which is possible if diversity is stable or if different strains employ different resources. There are at least tens of thousands of bacterial phylotypes in the gut, exact diversity characterizations being limited by sequencing depth \cite{claesson2009comparative}. Among all these types, some are much more abundant than others, and multiple rare types exist. Let us consider an active population size of order $N=10^{7}$ bacteria. 

Let us now discuss the range of $\delta$ values. Estimating the full distribution of fitness effects of mutations is difficult~\cite{Bataillon14}, and many studies focus on beneficial mutations~\cite{Levy15} because of their importance in adaptation.
A recent microfluidic mutation accumulation experiment \cite{robert2018mutation} performed in \textit{Escherichia coli} allowed to suppress the effect of selection and to study all spontaneous mutations. The average fitness cost $\langle\delta\rangle$ was found to be $3 \times 10^{-3}$, but it largely arises from a minority of strongly deleterious mutations. The skewness of the distribution of fitness costs was found to be positive and large, of about 17. Accurate measurements of extremely small fitness effects are highly challenging, but a Gamma distribution of the same mean and skewness would yield 10\% of all mutations with a fitness cost $\delta$ smaller than $10^{-7}$. Thus, a substantial fraction of all spontaneous mutations occurring in gut bacteria is expected to be effectively neutral.

\section{Validation by stochastic simulations}
\label{sec:stochsim}

\subsection{Stochastic simulation methods}

In our stochastic simulations, space and time are discretized. We denote the discrete steps in space and time used in these stochastic simulations by $\delta x$ and $\delta t$. Recall that our model is one-dimensional and therefore the only spatial dimension to be considered is $x$, along the gut main axis. To simplify notations, in this section we denote by $F(x)$ the linear concentration of food and by $B(x)$ that of bacteria (which amounts to taking a gut section with unit area). 

\paragraph{Discretizing transport. }
To represent transport in our discrete simulations, rates of exchange of food and bacteria between adjacent segments of lengths $\delta x$ are defined. 

First, to represent diffusive flow, let us introduce the rate $r_d$ at which food moves from one segment to each one of its two closest neighboring segments (upstream or downstream). For the discrete model to converge to our continuous equations as $\delta x\to 0$, $r_d$ has to satisfy:
\begin{equation}
\lim_{\delta x \to 0} r_d\,\delta x\left[ F(x+\delta x) - F(x) \right]= D  \frac{\partial F}{\partial x}\,,
\end{equation}
where we expressed the diffusive flow per unit time from the segment at $x$ to its immediate downstream neighbor both in the discrete and in the continuous descriptions. 
This leads to:
\begin{equation}
r_d =\frac{D}{\delta x^2}\,.
\end{equation}

Similarly, to represent convective flow, let us define the rate $r_v$ at which food moves from one segment to its downstream neighbor. For the discrete model to converge to our continuous equations as $\delta x\to 0$, $r_v$ has to satisfy: 
\begin{equation}
\lim_{\delta x \to 0} r_v\,\delta x \,F(x) = v \, F(x)\,,
\end{equation}
where we expressed the convective flow per unit time from the segment at $x$ to its immediate downstream neighbor both in the discrete and in the continuous descriptions. 
This leads to:
\begin{equation}
r_v =\frac{v}{\delta x}\,.
\end{equation}

Then overall, in the discrete model, during $\delta t$, the amount of food moving from one segment to its upstream neighbor is $r_-\delta t F(x)\delta x$ where $r_-$ is the rate of upstream transport:
\begin{equation}
    r_-=r_d=\frac{D}{\delta x^2}\,.
    \label{rminus}
\end{equation}
Meanwhile, the amount of food moving from one segment to its downstream neighbor is $r_+\delta t F(x)\delta x$ where $r_+$ is the rate of upstream transport:
\begin{equation}
    r_+=r_d+r_v=\frac{D}{\delta x^2}+\frac{v}{\delta x}\,.
    \label{rplus}
\end{equation}
Note that the mapping with the continuous equations is obtained for $\delta x\to 0$, and thus results may deviate for finite $\delta x$. 

Similarly, during $\delta t$, individual bacteria have a probability $r_-\delta t$ to jump upstream, and $r_+\delta t$ to jump downstream.

\paragraph{Discretizing growth.}

Let us start from our continuous equations on concentrations of bacteria and food, and map them to a description with discrete bacteria. 
Recall that in this section, $B$ and $F$ are linear concentrations of bacteria and food. Restricting to the terms of Eq.~\ref{eq:PDEs} modeling growth, we have:
\begin{equation}
\begin{dcases}
\frac{dB}{dt}&= r B \frac{F}{F+k}\,,\\
 \frac{dF}{dt}&= - \frac{1}{\alpha} r B \frac{F}{F+k}\,.
 \end{dcases}
\end{equation}
Denoting by $b(x)=B(x)\delta x$ (resp. $f(x)=F(x)\delta x$) the average number of bacteria (resp. the quantity of food in moles) present in the segment $[x,x+\delta x]$, these equations become: 
\begin{equation}
\begin{dcases}
\frac{db}{dt}&= r b \frac{f}{f+\tilde{k}}\,,\\
 \frac{df}{dt}&= - \frac{1}{\alpha} r b \frac{f}{f+\tilde{k}}\,,
 \end{dcases}
 \label{fb}
\end{equation}
$\tilde{k}=k\delta x$. Similarly, let us introduce $f_{\text{in}}=\fin \delta x$.

Each individual bacterium has a reproduction rate of $r f(x)/[f(x)+\tilde{k}]$, so during $\delta t$ it has a probability $r \delta t\, f(x)/[f(x)+\tilde{k}]$ to reproduce. If one bacterium reproduces, then $f(x)$ is decreased by $1/\alpha$, thereby ensuring the mapping with Eqs.~\ref{fb}. Note that, since this quantity is fixed, for small values of $\delta x$, it may lead to transient negative local values of $f(x)$, due to discretization and to the assumption that food is only consumed in the segment where replication occurs. When this happens, we set the local reproduction rate of bacteria to zero, and thanks to diffusion, such unrealistic states are short-lived.

\paragraph{Simulation steps.}

\begin{itemize}

\item{Initialization:} The initial discrete numbers $n_b(x)$ of bacteria in each segment $[x,x+\delta x]$ are obtained from the steady-state numerical solution $B(x)$ of the continuous equations through $n_b(x)=\lceil b(x) \rceil = \lceil B(x) \delta x \rceil$, where $\lceil .\rceil$ denotes the ceiling function.

\item{Equilibration without mutants:}

\begin{enumerate}
 \item During $\delta t$, a bacterium in $[x,x+\delta x]$ has a probability $r f(x)/[f(x)+\tilde{k}]$ of reproducing. For each bacterium, a random number is drawn in a uniform distribution on the interval $[0,1]$, and compared to the reproduction probability. This determines whether this bacterium reproduces.
 \item For each bacterium within $[x,x+\delta x]$ that reproduces, the quantity of food $f(x)$ is decreased by $1/\alpha$.
 \item After the steps above, food is updated according to the following rule: $f(x) \leftarrow f(x) \left[ 1- (r_+ + r_-)\delta t \right] + f(x+\delta x) r_- \delta t +  f(x-\delta x) r_+ \delta t$, where the rates $r_-$ and $r_+$ are given respectively by Eq.~\ref{rminus} and Eq.~\ref{rplus}. On the boundaries,  $f(0) \leftarrow f(0) \left[ 1- (r_+ + r_-)\delta t \right] + f(\delta x) r_- \delta t +  \left[ v f_{in}/\delta x+r_- f(0) \right] \delta t$, and $f(L) \leftarrow f(L)\left( 1-r_+ \delta t \right) + f(L-\delta x)r_+ \delta t$.
 \item For each bacterium at each site $x$ a random number is drawn from a uniform distribution on an interval $[0,1]$, and compared to the probability to move to the right, $r_+ \delta t$, to move to the left, $r_- \delta t$, or to stay, $1-\left( r_++r_- \right)\delta t$. Bacteria move accordingly.
 \end{enumerate}

The above steps 1 to 4 are repeated until time $t$ reaches $1000$ hours.
A new initial state of bacteria and food is obtained by taking the average in the time interval $t\in[100,1000]$~h. Then, steps 1 to 4 are repeated for $100$ more hours.

\item{Mutation:} At the next replication of one bacterium, we assign to it one mutant daughter bacterium. If more than one bacterium replicates during $\delta t$, the mutant daughter bacterium is assigned to one of them, chosen uniformly at random.

\item{Evolution to fixation or extinction:} The above steps 1 to 4 are repeated until extinction or fixation of the mutant bacteria, keeping track of what bacteria are mutant and wild-type. 141,681 different stochastic replicates were run.

\end{itemize}

\begin{figure}[h!]
\includegraphics[width=\textwidth]{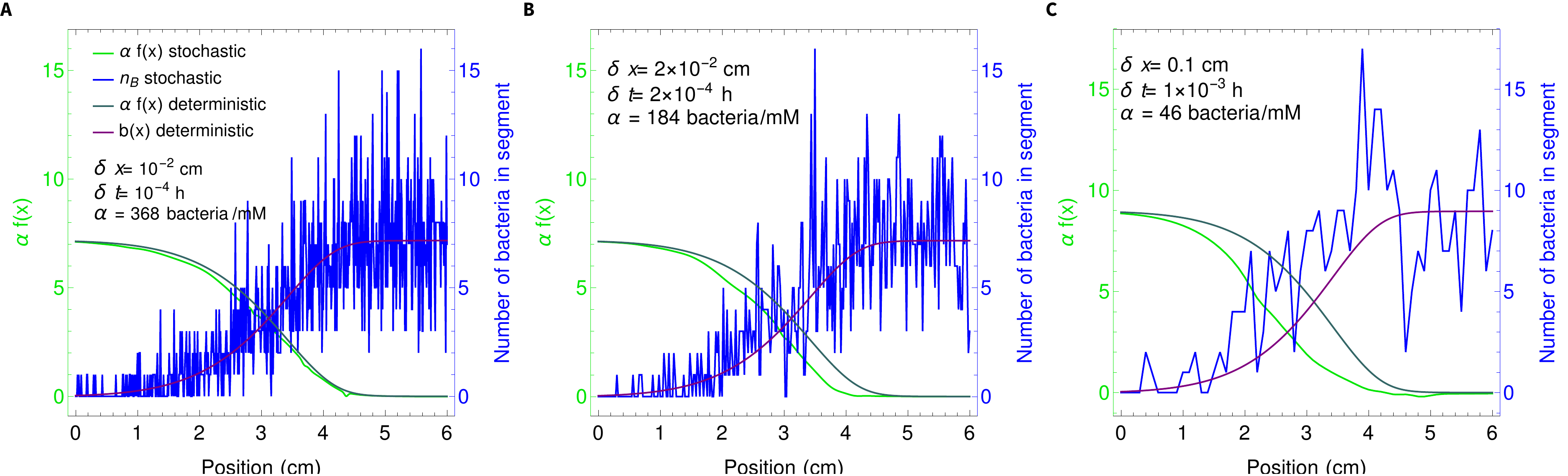}
\caption{\textbf{Dependence of the stochastic profiles on the discretization parameters, $\delta x$, $\delta t$, and $N_\text{T}$ (via $\alpha$).} Numbers of bacteria and rescaled food quantities at time $10^3$ h are shown in the stochastic and in the deterministic description versus position $x$ in the gut. Initial conditions are the same in all panels and are taken as described in the ``Initialization'' item of our simulation steps. The only difference between panels is the value of the discretization parameters.
Other parameters of the system are as in Fig.~\ref{fig1}B, apart from $v=0.514$ cm/h (and $\alpha$ which is specified in each panel).
}\label{fig:StochSimDiscr}
\end{figure}

\paragraph{Choice of the discretization parameters.} Three parameters describe the discrete nature of this stochastic simulation: the discrete spatial step $\delta x$, the discrete time step $\delta t$, and the finite total number of bacteria $N_\text{T}=\sum_x b(x)=\int_0^L B(x) dx $. In the numerical resolutions of the main text, we focused on large bacterial population sizes. But here, as we aim to validate the results from our deterministic model using a stochastic one, we consider less realistic small population sizes, which allows for shorter computation times. However, $N_\text{T}$ should not be too small in order to avoid spontaneous extinctions of bacteria due to fluctuations within the time of the simulations. In practice, to control population size, we tune the conversion factor $\alpha$ between food and bacteria. In addition, taking $\delta x$ or $\delta t$ not small enough leads to shifts in the food profile compared to the continuous case, as demonstrated in Fig.~\ref{fig:StochSimDiscr}.

In order to keep simulation times reasonably short, we chose the parameters corresponding to Fig.~\ref{fig:StochSimDiscr}B. Note that this case exhibits a small shift in the spatial profile compared to the continuous case. However, this shift is stable, the steady state relation $B=\alpha(F_{in}-F)$ holds on average, and the source of the shift with respect to the continuous case is known to be the choice of $\delta x$ and $\delta t$. The fluctuations caused by the choice of $N_\text{T}$ (via $\alpha$) in Fig.~\ref{fig:StochSimDiscr}B are also stable and are not causing  extinctions or changes of spatial regimes.

\subsection{Stochastic simulation results}

Here, we present results obtained over 141,681 different stochastic replicates of the simulation described above, with discretization parameters corresponding to Fig.~\ref{fig:StochSimDiscr}B.

\begin{figure}[h!]
\includegraphics[width=\textwidth]{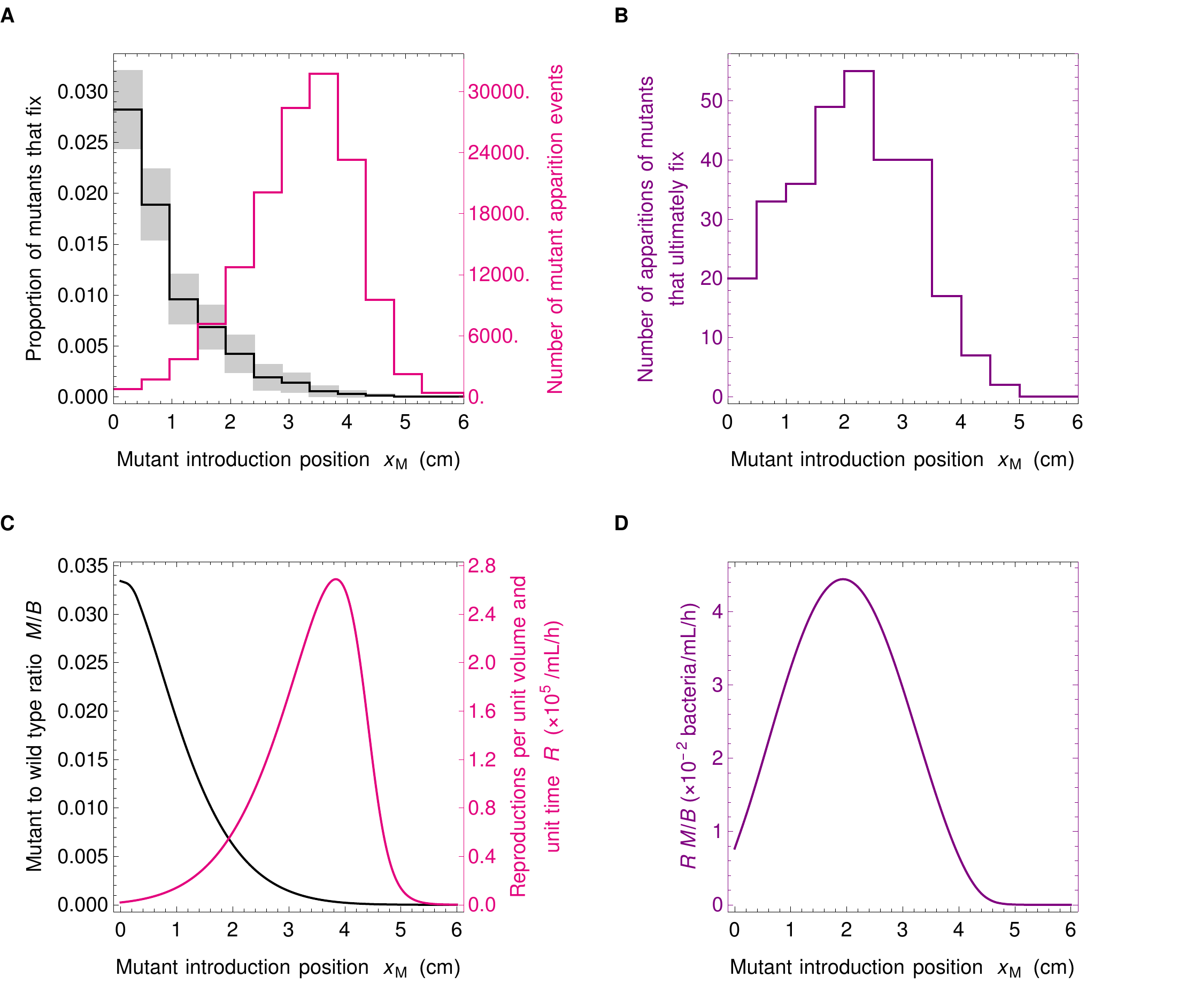}
\caption{
\textbf{Fate of neutral mutants appearing at various locations in the gut: stochastic and deterministic descriptions.} 
\textbf{A:} Black: proportion $p(\xM)$ of mutant bacteria that fix versus their introduction position $\xM$, computed as $p(\xM)=N_{MF}(\xM)/N_{MA}(\xM)$, where $N_{MF}(\xM)$ is the number of mutants that fix in the segment at $\xM$, while $N_{MA}(\xM)$ is the number of mutants that appeared in that bin. Gray shaded boxes correspond to $p(\xM)\pm \sqrt{N_{MF}(\xM)}/{N_{MA}(\xM)}$. Pink: histogram of the number of mutant apparition events at each introduction position $\xM$.
\textbf{B:} Histogram of the number of apparition events of mutant bacteria that eventually fix versus their introduction position $\xM$. In panels A and B, parameter values are as in Fig.~\ref{fig1}B, except $\alpha=$ \SI[per-mode = symbol]{184}{\bacteria \per\milli\Molar} and  $v=$ \SI[per-mode = symbol]{0.514}{\centi\meter\per\hour}. Discretization parameters are $\delta x=$\SI[per-mode = symbol]{0.02}{\centi\meter}, $\delta t=$\SI[per-mode = symbol]{2e-4}{\hour}.  The bin size used is 0.5 cm.
\textbf{C} and \textbf{D:} Corresponding numerical solutions of the deterministic system. In panels C and D, parameter values are as in Fig.~\ref{fig1}B, except $v=$ \SI[per-mode = symbol]{0.514}{\centi\meter\per\hour}. For direct comparison with the stochastic results shown in panels A and B, curves have been rescaled, the black one by $\alpha_d/(M_{0,d} \alpha_s)$, the pink one by $\alpha_s/\alpha_d$, and the purple one by $1/M_{0,d}$, where $\alpha_d$ is the value of $\alpha$ used in the deterministic case (see Fig.~\ref{fig1}B), while $\alpha_s$ is the one employed in the stochastic case, and $M_{0,d}$ is the initial local mutant concentration at $\xM$ in the deterministic description. 
 }\label{fig:fig2_stoch}
\end{figure}

Fig.~\ref{fig:fig2_stoch} shows the fate of neutral mutants appearing at various locations in the gut. We observe a good agreement between panels A and B, obtained from our stochastic simulations, and panels C and D, obtained from our numerical resolution of the deterministic system (analogous to Fig.~\ref{fig2}). In particular, the black curve in Fig.~\ref{fig:fig2_stoch}A, showing the proportion $p(\xM)$ of mutant bacteria that fix versus their introduction position $\xM$, matches the black curve in Fig.~\ref{fig:fig2_stoch}C, which shows the steady-state ratio $M/B$ of mutant to wild-type bacteria concentrations. This corroborates the fact that the deterministic steady-state ratio $M/B$ in the deterministic description behaves as the mutant fixation probability in the stochastic case, for each given mutant introduction position $\xM$.

\newpage

\begin{figure}[h!]
\includegraphics[width=.6\textwidth]{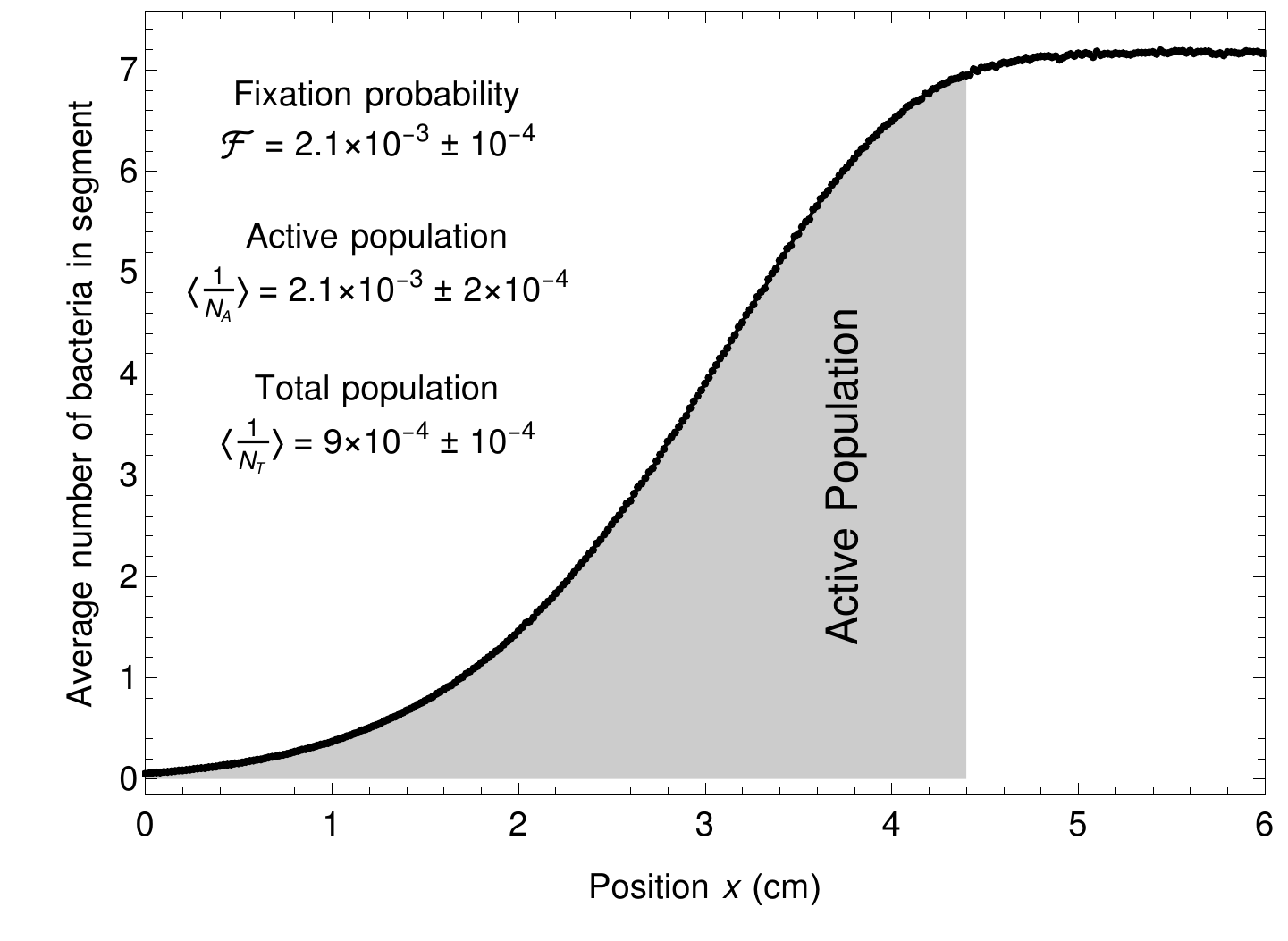}
\caption{\textbf{Fixation probability is equal to the inverse active population size.} The curve shows the average profile of the bacterial population in the stochastic simulations run in panels A and B of Fig.~\ref{fig:fig2_stoch}. The total population size is proportional to the total area below this curve, while the active population size (defined in section~\ref{SM:active_pop}, see also Fig.~\ref{fig4} of main text) is proportional to the area shaded in gray. The fixation probability $\mathcal{F}$ is computed from our stochastic simulation results as $\mathcal{F}=N_F/N_R$, where $N_F$ is the number of fixation events and $N_R$ the number of runs. The associated confidence interval is given by $(N_F \pm \sqrt{N_F})/N_R$. The inverses of the total and of the active population size are also given, averaged over realizations and over time. The associated confidence intervals are estimated via standard deviations.
}\label{fig:stoc_AP}
\end{figure}

One of the main results of our study is that the fixation probability of a mutant is given by $1/N_\text{A}$ instead of $1/N_\text{T}$. Fig.~\ref{fig:stoc_AP} demonstrates that this result is validated by our numerical simulations. 

\newpage

\begin{figure}[h!]
\includegraphics[width=\textwidth]{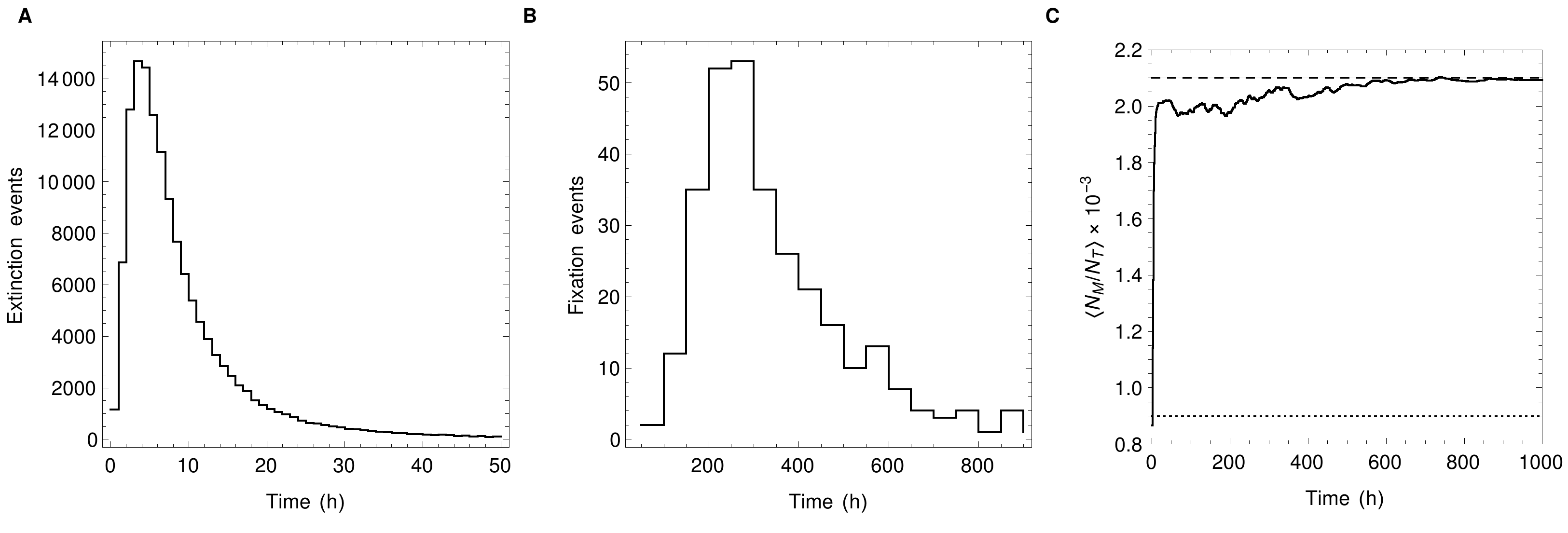}
\caption{  \textbf{Time scales in stochastic simulations.} \textbf{A:} Histogram of mutant extinction times. \textbf{B:} Histogram of mutant fixation times. \textbf{C:} Curve: average proportion of mutants versus time after mutant apparition. Dashed line: $\langle 1/ N_\text{A} \rangle$. Dotted line: $\langle 1/ N_\text{T} \rangle$. Upon mutant apparition, as there is one mutant, the mutant proportion is $1/N_\text{T}$. In the long term, as the population will be either all wild-type or all mutant, the average mutant proportion tends to the fixation probability. But the time scale over which the mutant proportion becomes close to $1/N_\text{A}$ is small compared to the time to fixation.   
}\label{fig:stoc_times}
\end{figure}

Furthermore, Fig.~\ref{fig:stoc_times}A-B shows the distribution of mutant extinction times and mutant fixation times. Most extinction times are relatively short. The peak of the distribution of the position of mutant apparition (pink curve of panel A of Fig.~\ref{fig:fig2_stoch}) is about 2~cm from the downstream boundary, which takes about $4$~h to cross at the velocity $v$ employed here. Consistently, this is also the peak time for extinction events. Fixation times are much longer. Since the effective population size is the active population size, of the order of 470 and the replication rate is $r=0.42\,\,\textrm{h}^{-1}$, the fixation time is expected to be of the order of $N_\text{A}/r \simeq 1000$~h, which is indeed the right order of magnitude. As bacterial populations in the gut are very large, fixation times will accordingly be very large. Fig.~\ref{fig:stoc_times}C shows the average proportion of mutants versus time. At the time of mutant apparition, it is the inverse of the total population size, but very quickly this mutant proportion becomes close to the inverse of the active population size. Therefore, the effect of active versus total population size matters much earlier than the fixation time.

\end{document}